\documentclass[12pt]{article}

\usepackage{graphicx,amsmath,amssymb,array,calc}
\usepackage{bm}
\usepackage{times}

\newcommand{\bea}{\begin{eqnarray}}
\newcommand{\eea}{\end{eqnarray}}
\newcommand{\be}{\begin{equation}}
\newcommand{\ee}{\end{equation}}
\newcommand{\nn}{\nonumber}
\usepackage{graphicx,fullpage}
\usepackage{amsmath,amssymb,array,calc}

\def\k{{\bm k}}

\long\def\symbolfootnote[#1]#2{\begingroup%
\def\thefootnote{\fnsymbol{footnote}}\footnote[#1]{#2}\endgroup} 

\begin {document}

\setlength\textheight{9in}
\setlength\textwidth{7.5in}
\begin{center}
   {\Large\bf Second order hydrodynamic coefficients from 3-point stress tensor correlators via AdS/CFT}
\vskip 0.5cm

{\large Peter Arnold, Diana Vaman, Chaolun Wu, Wei Xiao}\symbolfootnote[1]{E-mail addresses: parnold, dv3h, cw2an, wx2m@virginia.edu}
\vskip 0.5cm

\it {Department of Physics, University of Virginia\\
Box 400714, Charlottesville, Virginia 22904, USA}\\ 
\end{center}
\begin{abstract}
We study second order relativistic viscous hydrodynamics in 4-dimensional conformal field theories.  We derive Kubo-type relations for second order hydrodynamic coefficients in terms of 3-point stress tensor retarded correlators. For ${\cal N}{=}4$ super Yang-Mills theory at strong coupling and at finite temperature we compute these stress tensor 3-point correlators, using AdS/CFT, by evaluating {\it real-time} cubic  Witten diagrams in the AdS-Schwarzschild background.  The small momentum expansion of the 3-point correlators in terms of first and second order hydrodynamic coefficients is matched with the AdS result. We arrive at the same expressions for the hydrodynamic coefficients which multiply terms quadratic in the shear and vorticity tensors in the hydrodynamic expansion of the stress tensor as did Bhattacharyya, Hubeny, Minwalla and Rangamani \cite{Bhattacharyya:2008jc}. Our method extends the results of Baier et al \cite{Baier:2007ix}, and allows for a unified treatment of hydrodynamic coefficients, which are extracted from 2-, and now, 3-point retarded stress tensor correlators in the AdS-Schwarzschild background.
\end{abstract}
\newpage 
\section{Introduction and Summary}

One of the most familiar applications of AdS/CFT \cite{Maldacena:1997re}, \cite{Witten:1998qj} is computing linearized hydrodynamics for a variety of gauge theories with holographic duals (see \cite{ss} and references therein). More recently, second order hydrodynamic coefficients have been extracted using two different methods. Baier et al \cite{Baier:2007ix} used Kubo-like formulae for 2-point stress tensor correlators  to access a certain subset of second-order hydro coefficients of ${\cal N}{=}4$ super Yang-Mills plasma. Their computation was made possible by a real-time finite temperature prescription for computing 2-point correlators \cite{Son:2002sd, Herzog:2002pc}. On the other hand, Bhattacharya et al \cite{Bhattacharyya:2008jc} wrote a generalized black brane metric ansatz, dependent on the temperature and the black branes velocity viewed as collective fields, and turned the Einstein equations of motion solved perturbatively in the collective modes near the AdS boundary into equations of fluid dynamics. Specifically, the gravitational stress tensor, expanded near the boundary of AdS, took the form of a  non-linear fluid dynamics stress tensor, with the various terms in the expansion in velocity and gradients multiplied by the corresponding hydrodynamic coefficients. The subsets of hydro coefficients computed by  \cite{Baier:2007ix} and \cite{Bhattacharyya:2008jc} agreed where they overlapped, and between them determined the full set of coefficients. By now, the second order hydrodynamic coefficients have been computed following \cite{Bhattacharyya:2008jc} in a variety of cases: at finite chemical potential \cite{Erdmenger:2008rm, Banerjee:2008th}, or in the presence of fundamental matter \cite{Bigazzi:2009tc}. For recent review papers see 
\cite{Hubeny:2010ry} and \cite{CasalderreySolana:2011us}. The breakdown of second order hydrodynamics is investigated in \cite{Kovtun:2011np}.

Our work is intended as a continuation of \cite{Baier:2007ix}, where Kubo-like formulae are used in conjunction with higher-order stress tensor correlators to extract the second order hydro coefficients computed by \cite{Bhattacharyya:2008jc}. We first derive the necessary Kubo relations using the method of Moore and Sohrabi \cite{Moore:2010bu}\footnote{Readers should be warned that there is an error in the derivation of the Kubo relations for $\lambda_1,\lambda_2$ and $\lambda_3$ in \cite{Moore:2010bu}. Specifically, they left out the $(\epsilon+P) U^x U^x + P g^{xx}$ term in deriving their $T^{xx}$. Also, $U^i=O(h^2)$  only in the static limit.  }, then compute 3-point retarded stress tensor correlators in the hydrodynamic regime (i.e. in the limit of small momenta) in the AdS-Schwarzschild background which is dual to finite-temperature strongly coupled ${\cal N}{=}4$ super Yang-Mills in the limit of large number of colors $N_c\gg1$. In other words, as opposed to \cite{Bhattacharyya:2008jc}, we do not deform the gravitational background, but instead compute higher-order correlators in the background. The problem of computing such higher-order real-time finite temperature correlators was solved in \cite{Barnes:2010jp}. For momentum-space retarded 3-point stress tensor correlators this amounts to computing real-time Witten diagrams, depicted in Fig. 1b, with three causal (two advanced and one retarded) graviton bulk-to-boundary propagators, joined at a bulk vertex which is integrated up to the black hole horizon (i.e. in the maximal causal diamond)\footnote{See also \cite{CaronHuot:2009iq} for similar causality considerations and \cite{vanRees:2009rw} for a different take on this subject.}.
 
\begin{figure}
\begin{center}
\includegraphics[width=4.5in]{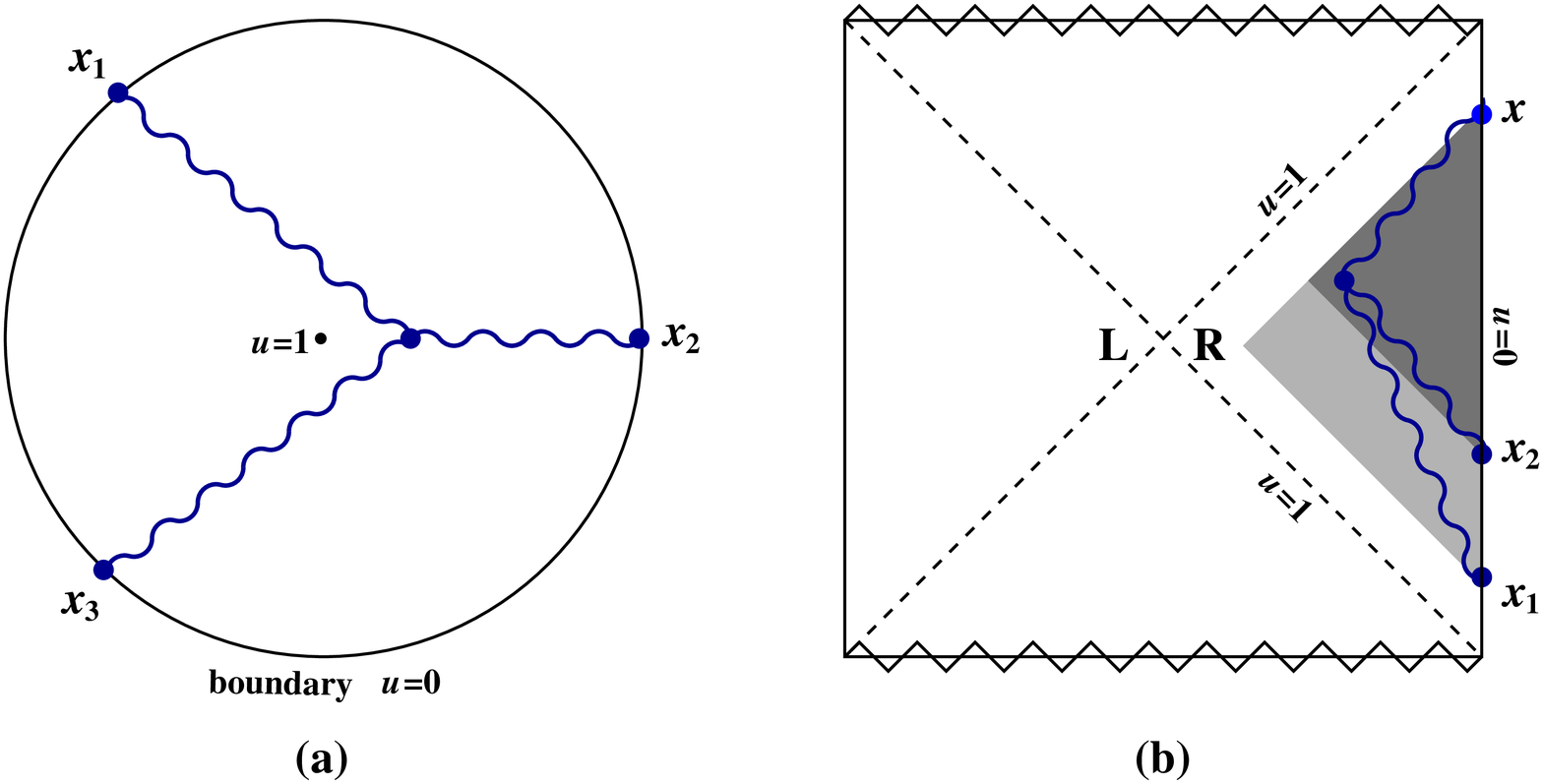} \\
Figure 1: Witten diagram for (a) 3-point correlator in imaginary time AdS-Schwarzschild and (b) retarded 3-point correlator of type ${\rm raa}$ with the boundary point $\bm x$ having the largest time; $\bm x_1$ and $\bm x_2$ can have any time order.
\end{center}
\end{figure}

To set up the problem, let us begin by revisiting some of the definitions and results of \cite{Baier:2007ix,Moore:2010bu}.

The stress tensor of a conformal fluid can be written in terms of an equilibrium piece plus an extra term, which in the hydrodynamic regime can be series-expanded in gradients:
\bea
&&T^{\mu\nu}=T^{\mu\nu}_{\text{eq}}+\Pi^{\mu\nu}, \qquad
T^{\mu\nu}_{\text eq}=(\epsilon+P)U^\mu U^\nu + P g^{\mu\nu} \label {eq:T}
\\
&&\Pi^{\mu\nu}=-\eta \sigma^{\mu\nu}+\eta \tau_{\Pi}\bigg( {}^{\langle}U\cdot
\nabla\sigma^{\mu\nu\rangle}+\tfrac 13 (\nabla\cdot U) \sigma^{\mu\nu}\bigg)
+\kappa\bigg(R^{\langle \mu\nu\rangle }-2 U_\rho U_\sigma R^{\rho\langle\mu\nu
\rangle\sigma}\bigg)\nn\\
&&\qquad+
\lambda_1 \sigma^{\langle \mu}{}_\rho \sigma^{\nu\rangle \rho}+
\lambda_2 \sigma^{\langle \mu}{}_\rho \Omega^{\nu\rangle \rho}+
\lambda_3 \Omega^{\langle \mu}{}_\rho \Omega^{\nu\rangle \rho}+\dots\label{eq PI}
\eea
Here $U^\mu$ is the fluid's velocity, normalized such that 
\be
U^\mu U^\nu g_{\mu\nu}=-1,
\ee
and $\Pi^{\mu\nu}$ is defined such that $U_\mu \Pi^{\mu\nu}=0$.
 $g_{\mu\nu}$ is the background metric, whose curvature tensor is $R_{\rho\mu\nu}{}^{\sigma}$, and all derivatives are covariant with respect to the background metric. $\sigma^{\mu\nu}$, $\Omega^{\mu\nu}$ are the shear and vorticity tensors respectively:
\bea
&&\sigma^{\mu\nu}=2\nabla^{\langle \mu} U^{\nu \rangle}\equiv
\Delta^{\mu\rho}\Delta^{\nu\sigma}(\nabla_\rho U_\sigma+
\nabla_\sigma U_\rho)-\tfrac 23 \Delta^{\mu\nu}\Delta^{\rho\sigma}\nabla_\rho U_\sigma \label{shear} \\
&&\Omega^{\mu\nu}=\tfrac 12 \Delta^{\mu\rho}\Delta^{\nu\sigma}(\nabla_\rho U_\sigma-\nabla_\sigma U_\rho)
\eea
where $\Delta^{\mu\nu}$ are transverse (to the fluid's velocity) projectors:
\be
\Delta^{\mu\nu}=g^{\mu\nu}+U^\mu U^\nu.
\ee
Angular brackets denote transverse projection, followed by symmetrization and removal of a trace:
\be
O^{\langle \mu\nu \rangle}= \tfrac 12 \Delta^{\mu\rho}\Delta^{\nu\sigma}(O_{\rho\sigma}+O_{\sigma\rho})-\tfrac 13 \Delta^{\mu\nu}\Delta^{\rho\sigma}O_{\rho\sigma}
\ee
 For a conformal fluid, energy density and pressure are related by the condition that the stress tensor is traceless, and so $\epsilon=(d-1)P$, where $d$ is the number of space-time dimensions.

In \cite{Moore:2010bu}, the fluid's response to a small metric perturbation was computed by solving the stress tensor conservation law
\be
\nabla_\mu T^{\mu\nu}=0,\label{cons law}
\ee
together with the condition that the fluid describes a conformal theory,
\be
T^\mu_\mu=0 \label{conf},
\ee
order-by-order in a double series expansion in the metric fluctuation and in gradients. This expansion of the stress tensor is then compared with
\bea
\langle T^{\mu\nu}(\bm z)\rangle_h&=&\langle T^{\mu\nu}\rangle_{h=0}-\tfrac 12
\int d^4 \bm x \,G^{\mu\nu| \rho\sigma}_{\rm ra}(\bm z;\bm x)h_{\rho\sigma}(\bm x)
\nn\\
&+&
\tfrac 18\int d^4 \bm x\int d^4 \bm y\, G^{\mu\nu| \rho\sigma| \tau\zeta}_{\rm raa}(\bm z;\bm x,\bm y)
h_{\rho\sigma}(\bm x) h_{\tau\zeta}(\bm y)+\dots\label{exp correl}
\eea
where $G^{\mu\nu|\dots}_{\rm ra\cdots a}$ are retarded $n$-point correlators, with the  measurement point $\bm z$ having the largest time. We assume the metric fluctuations $h_{\rho\sigma}$ vanish in the far past. To avoid clutter, for the rest of the paper we will suppress the ${\rm ra, raa}$ subscripts signifying retarded correlators\footnote{ For a nice summary of the ${\rm (r,a)}$ notation
 see \cite{Wang:1998wg}.} with the understanding that all the correlators we compute are of this type.

By identifying the two series expansions, namely the solution to (\ref{cons law},\ref{conf}) and  the expansion in retarded correlators (\ref{exp correl}), one gains access to the hydrodynamic expansion coefficients in terms of causal stress tensor correlators. For example, the shear viscosity, $\eta$, can be computed from 2-point correlators expanded to linear order in gradients \cite{Policastro:2002se}. Similarly, $\eta \tau_\Pi$ and $\kappa$ can also be computed from 2-point correlators, expanded to second order in gradients \cite{Baier:2007ix}. $\lambda_1, \lambda_2$ and $\lambda_3$ require 3-point correlators, expanded to second order in gradients \cite{Moore:2010bu}.

Since solving (\ref{cons law}) is easier in momentum space, we will write the solution to (\ref{cons law}) and (\ref{conf}) as a series expansion in momenta. For a direct comparison with the response of the stress tensor we need then the Fourier transform of (\ref{exp correl}):
\bea
\langle T^{\mu\nu}(\bm q)\rangle_h&=&\langle T^{\mu\nu}\rangle_{h=0}
-\tfrac 12 \int d^4\bm q_1 \frac{1}{(2\pi)^4}
\delta^4(\bm q-\bm q_1) G^{\mu\nu|\rho\sigma}(\bm q;-\bm q_1)h_{\rho\sigma}(\bm q_1)\nn\\
&+&\tfrac 18 \int d^4\bm q_1 \int d^4\bm q_2 
\frac{1}{(2\pi)^4}\delta^4(\bm q-\bm q_1-\bm q_2) G^{\mu\nu|\rho\sigma|\tau\zeta}(\bm q;-\bm q_1,-\bm q_2)h_{\rho\sigma}(\bm q_1)h_{\tau\zeta}(\bm q_2)+\dots
\nn\\ \label{exp correl FT}
\eea
where we have used translation invariance to factor out momentum conservation delta-functions.
In a 3-point function, there are two independent momenta,
and the spatial momenta could point in different directions.  However, for simplicity, we will assume that they
{\it both}\/ point in the $z$ direction (i.e. we consider  metric fluctuations which are independent of the $x,y$ coordinates): 
\be
\bm q_1^\mu=(\omega_1,0,0,k_1), \bm q_2^\mu=(\omega_2,0,0,k_2).
\ee
The purpose of this paper is to obtain the remaining second order hydrodynamic coefficients $\lambda_1,\lambda_2$ and $\lambda_3$ from 3-point stress tensor correlators. 
We will find it simplest to study the $xy$-component $\langle T^{xy}\rangle_h$
and will extract second-order hydrodynamic coefficients from $G^{xy| ..| ..}$:
\bea
&&\lim_{\substack{\omega_1\to0 \\ \omega_2\to 0}}
\partial_{\omega_1}\partial_{\omega_2} \lim_{\substack{k_1\to0 \\ k_2\to 0}}G^{xy| xz| yz}=-\lambda_1+\eta\tau_\Pi\nn\\
&&\lim_{\substack{\omega_1\to 0\\k_2\to 0}}\partial_{k_2}\partial_{\omega_1}\lim_{\substack{\omega_2\to 0\\k_1\to 0}}G^{xy| yz| 0x}=-\tfrac 14\lambda_2+\tfrac 12 \eta\tau_{\Pi}\nn\\
&&\lim_{\substack{k_1\to0\\k_2\to0}}
     \partial_{k_1} \partial_{k_2}
      \lim_{\substack{\omega_1\to0\\\omega_2\to0}}
            G^{xy| 0x| 0y}
   =
   -\tfrac14 \lambda_3 .\label{preview}
\eea
We will find that the leading order AdS/CFT computation of the correlators leads to the same expressions for $\lambda_1, \lambda_2, \lambda_3$ as those previously obtained by \cite{Bhattacharyya:2008jc}:
\be
\lambda_1=\frac{N_c^2 T^2}{16}, \qquad \lambda_2 = -\frac{N_c^2 T^2 \ln 2}{8}, \qquad \lambda_3=0.\label{preview2}
\ee

The paper is organized as follows. In Section 2 we derive the Kubo relations given in (\ref{preview}). More Kubo formulae can be found in Appendix \ref{kubo}. In Section 3 we use AdS/CFT to evaluate the stress tensor correlators. We begin with a review of 2-point stress tensor correlators, and we comment how the hydrodynamic expansion of the retarded correlators and the AdS/CFT expressions are matched term-by-term in a gradient (small momenta) expansion. Then we present a similar result for the 3-point retarded stress tensor correlators which appear in (\ref{preview}), with the final result for the second-order hydro coefficients given by (\ref{preview2}). Technical details are relegated to appendices. For example, the $\eta\tau_\Pi$ contribution to (\ref{preview}) is derived in Appendix A.  The graviton bulk-to-boundary propagators, expanded up to third order in momenta, are given in Appendix C. A few 2-point correlators are discussed in the text, but the rest of them are presented in Appendix D. Lastly, the on-shell (first, second and third order) gravitational action is given in Appendix E.

We use the following conventions:  five-dimensional tensors, which live in the AdS-Schwarzschild geometry, have indices given by $K,L,M,\dots=0,1,2,3,5=t,x,y,z,u$; 
 four-dimensional space-time indices are $\mu,\nu,\dots=0,1,2,3$; three-dimensional spatial indices are $i, j, \dots=1,2,3=x,y,z$. 4-vectors are denoted by
bold letters $(\bm x,\bm q)$, and 3-vectors with an arrow ($\vec x$). 
We use a bar to denote background values, in both the field theory and in the holographic dual. E.g. $\bar \epsilon$ is the field theory background energy density, and $\bar g_{MN}$ denotes the background AdS-Schwarzschild metric.
For simplicity of notation, for the rest of the paper we will also work in units where $2\pi T=1$.

\section{Kubo formulae for $\lambda_1,\lambda_2$ and $\lambda_3$}

\subsection{The fluid velocity to leading order in the metric fluctuations and gradients}

We begin by assuming that the fluid is initially in equilibrium in a flat space-time background, which is briefly distorted by a small gravitational perturbation:
\be
g_{\mu\nu}=\eta_{\mu\nu}+h_{\mu\nu}, \qquad \epsilon=\bar\epsilon+\delta \epsilon, \qquad P=\bar P+\delta P
\ee
where $h_{\mu\nu}(t,z)$ is a small metric fluctuation, and $\delta \epsilon$, $\delta P$ are the induced energy density and pressure variations. In this section we will keep things general, and we will not enforce $\bar P=\tfrac 13 \bar\epsilon$ until the end. We take the fluid to be initially at rest
\be
U=\bar U+\delta U,\qquad \bar U^\mu = (1,\vec 0),
\ee
and so the background transverse projectors are simply
\be
\bar\Delta^{ij}=\delta^{ij}, \qquad \bar\Delta^{0i}=\bar\Delta^{00}=0.
\ee

From the normalization condition $U^\mu U_\mu=-1$ we get the temporal component
\be
U^0=1+\tfrac 12 h_{00}+{O}(h^2).
\ee 
The spatial components $U^i$ are determined from the stress tensor conservation law:
\begin {equation}
   0 = \nabla_\mu T^{\mu\nu} =
   [\partial_\mu (\epsilon+P)] U^\mu U^\nu
   + (\partial_\mu P) g^{\mu\nu}
   + (\epsilon+P) \nabla_\mu (U^\mu U^\nu)
   + \nabla_\mu \Pi^{\mu\nu},
\label {eq:conserve1}
\end {equation}
where
\begin {equation}
   \nabla_\mu(U^\mu U^\nu) =
   \partial_\mu (U^\mu U^\nu) + \Gamma^\mu_{\mu\sigma} U^\sigma U^\nu
                            + \Gamma^\nu_{\mu\tau} U^\mu U^\tau.
\label {eq:simpuu}
\end {equation}
To leading order in the metric fluctuations, this can be simplified using
\begin {equation}
   [\partial_\mu (\epsilon+P)] U^\mu U^\nu
   + (\partial_\mu P) g^{\mu\nu}
   =
   \partial_\mu (\epsilon+P) \bar U^\mu \bar U^\nu
   + (\partial_\mu P) \eta^{\mu\nu}
   + O(h^2) 
\label {eq:simp1}
\end {equation}
by noting that
$\partial\epsilon$ and $\partial P$ are $O(h)$.

First consider the $\nu=\perp\equiv x,y$ case and keep only terms through
$O(h)$ in (\ref{eq:conserve1}).  Since $\bar U^\perp = 0$ and
$\partial_\perp = 0$, the first two terms in
(\ref{eq:conserve1}) vanish at $O(h)$ by (\ref{eq:simp1}).
Using (\ref{eq:simpuu}), and noting that both $U^\perp$ and $\Gamma$
are $O(h)$,
\begin {align}
   \nabla_\mu(U^\mu U^\perp) &=
   \partial_\mu (\bar U^\mu U^\perp)
       + {\Gamma^\perp}_{\mu\tau} \bar U^\mu \bar U^\tau
       + O(h^2)
\nonumber\\
   &= {\dot U}^\perp + \dot h_{0\perp} + O(h^2) ,
\end {align}
where we use dots as shorthand for time derivatives.
Now use the fact that
\begin {equation}
   U_\perp = g_{\perp\nu} U^{\nu}
   = g_{\perp 0} U^0 + g_{\perp\perp} U^\perp + O(h^2)
   = h_{\perp 0} + U^\perp + O(h^2)
\label {eq:uperplower}
\end {equation}
to rewrite this as
\begin {equation}
   \nabla_\mu(U^\mu U^\perp)
   = {\dot U}_\perp + O(h^2) .
\end {equation}

Putting it all together (along with the fact that $\Pi^{\mu\nu}$ is
$O(h)$ and so $\nabla\Pi \simeq \partial\Pi$),
the transverse case of the conservation
law (\ref{eq:conserve1}) is
\begin {equation}
   (\bar\epsilon+\bar P) {\dot U}_\perp
       + \partial_\mu \Pi^{\mu\perp}
   = O(h^2) .
\end {equation}
This simplifies to
\begin {equation}
   (\bar\epsilon+\bar P) {\dot U}_\perp
       + \partial_z \Pi^{z\perp}
   = O(h^2) ,
\label {eq:transverse}
\end {equation}
since $\Pi^{\mu\nu}$ has only spatial components at $O(h)$ and $\partial_\perp$
vanishes.

Our goal will be to work to leading order in frequencies
$\omega$ and leading order in spatial momenta
$k$, without assuming anything about the relative size of $\omega$
and $|k|$.  (That is, we want to be able to handle $\omega \sim k^2$
and $\omega \sim |k|$ and $\omega=0$ and $k=0$ on an equal footing.)
In (\ref{eq:transverse}), we have explicitly kept the term which is leading
order in time derivatives of $U$.  We  now need the term that is
leading order in spatial derivatives of $U$.  That comes from
the first-order hydro terms in $\partial_i \Pi^{i\perp}$.

So consider the first order hydro expansion of $\Pi^{\mu\nu}$:
\begin {align}
   \Pi^{\mu\nu}_{\rm 1st}
   &= -\eta \sigma^{\mu\nu}
     -\zeta \Delta^{\mu\nu} \Delta^{\alpha\beta} \nabla_\alpha U_\beta
\nonumber\\
   &= -\eta \Delta^{\mu\alpha} (\nabla_\alpha U_\beta + \nabla_\beta U_\alpha)
                         \Delta^{\beta\nu}
     + (\tfrac23\eta-\zeta)
            \Delta^{\mu\nu} \Delta^{\alpha\beta} \nabla_\alpha U_\beta
   ,
\end {align}
where $\zeta $ is the bulk viscosity and $\zeta=0$ for a conformal theory.
At first order in $h$,
\begin {align}
   \Pi^{ij}_{\rm 1st}
   &= -\eta (\nabla_i U_j + \nabla_j U_i)
     + (\tfrac23\eta-\zeta) \delta_{ij} \nabla_k U_k
     + O(h^2)
   ,
\end {align}
and in particular
\begin {align}
   \Pi^{z\perp}_{\rm 1st}
   &= -\eta (\partial_z U_\perp
                + \partial_0 h_{z\perp} - \partial_z h_{0\perp})
      + O(h^2) .
\label {eq:Pi1st}
\end {align}
Putting this together with (\ref{eq:transverse}), we get
\begin {equation}
   (\bar\epsilon+\bar P) {\dot U}_\perp - \eta U''_\perp
   - \eta\partial_z(\dot h_{z\perp} - h'_{0\perp})
   + \partial_z (\mbox{2nd order hydro effects})
   = O(h^2) ,
\end {equation}
where primes are shorthand for $z$-derivatives.
The solution is
\begin {equation}
   U_\perp =
   \frac{D_\eta k(k h_{0\perp} + \omega h_{z\perp})}
        {-i \omega + D_\eta k^2} 
   + O(\omega U_\perp, k U_\perp, h^2) ,
\label {eq:uperp}
\end {equation}
where
\begin {equation}
   D_\eta  \equiv \frac{\eta}{\bar\epsilon + \bar P}
\end {equation}
is the diffusion constant associated with shear viscosity.
Here $(\omega,0,0,k)$ is the four-momentum associated with
the factors of $h$.
The $O(\omega U_\perp, k U_\perp)$ corrections listed in
(\ref{eq:uperp}) represent corrections due to second-order hydro,
which are suppressed by additional factors of momenta\footnote{Our notation
$O(\omega U_\perp, k U_\perp)$ assumes that the leading order in $U_\perp$ is not zero.}.
Note that $U_\perp$ vanishes at $O(h)$ if one sets
$h_{\mu\perp} \propto q_\mu = (-\omega,0,0,k)$.

This leaves us with $U_z$, which we will determine by solving the remaining $\nu=z,0$  components of the stress tensor conservation law, to linear order in the metric fluctuation, and to leading order in gradients. First we compute
\be
\Pi^{zz}_{\rm 1st}\simeq-\tfrac 43\eta\bigg[U_z'+\tfrac12 (\dot h_{zz}-\tfrac 12\dot h_{\perp\perp}-2h'_{0z})\bigg]
-\zeta\bigg[U_z'+\tfrac 12 (\dot h_{ii}-2h'_{0z})\bigg],
\ee
where $h_{\perp\perp}\equiv h_{xx}+h_{yy}$. 
Then $\nabla_\mu T^{\mu z}=0$ yields
\be
\dot U_z-\bigg(\tfrac 43 D_\eta + \frac{\zeta}{\bar\epsilon+\bar P}\bigg)U_z''\simeq
-\frac{P'}{\bar \epsilon+\bar P}+\tfrac 12 h'_{00} +\tfrac 23 D_\eta (\dot{h}'_{zz}-\tfrac 12\dot h'_{\perp\perp}-2h_{0z}'')+\frac{\zeta}{2(\bar\epsilon+\bar P)}(\dot{h}'_{ii}-2h_{0z}''),\label{uz1}
\ee
while from $\nabla_\mu T^{\mu 0}=0$ we get
\be
U_z'\simeq h'_{0z}-\tfrac 12 \dot h_{ii}-\frac{\dot\epsilon}{\bar\epsilon+\bar P}\label{uz2}.
\ee

Specialize now to conformal theories ($\zeta=0, \bar P=\tfrac 13 \bar\epsilon$). If we stay away from the sound pole $\omega^2\simeq \tfrac 13 k^2$, then to leading order in derivatives we can ignore $\Pi^{zz}_{\rm 1st}$ and the $D_\eta$ terms in (\ref{uz1}), and combine (\ref{uz1}) and (\ref{uz2}) to get
\be
U_z\simeq\frac{1}{\partial_t^2-\tfrac 13 \partial_z^2}(\tfrac 12 \dot h'_{00}+\tfrac 16 \dot{h}'_{ii}-\tfrac 13 h_{0z}'').
\ee
and 
\be
\epsilon=\bar\epsilon+\tfrac 43 \frac{\bar\epsilon}{\partial_t^2-\tfrac 13 \partial_z^2}(\dot{h}'_{0z}-\tfrac 12 h_{00}''-\tfrac 12 \ddot{h}_{ii})+{\cal O}(h^2, \omega h, k h)\label{epsilon}.
\ee

In what follows we will derive Kubo relations for the second-order hydrodynamic coefficients from the response of the stress tensor component $T^{xy}$. From
(\ref{eq:T}) and (\ref{eq PI}),
\begin {equation}
   T^{xy} =
   \bar\epsilon \left[
       \tfrac43 U^{(1)x} U^{(1)y}
       + \tfrac13 (- h_{xy} + h_{x\mu} \eta^{\mu\nu} h_{\nu y})
   \right]
   + \tfrac13 (\delta^{(1)} \epsilon) (- h_{xy})
   + \Pi^{xy}
   + O(h^3) ,\label{txy}
\end {equation}
where the superscript $(1)$ denotes a contribution of order $h$ and $\epsilon=\bar\epsilon+\delta^{(1)}\epsilon+\cdots$.

\subsection{$\lambda_1$}

In the hydrodynamic expansion, $\lambda_1$ is the shear tensor-squared coefficient:
\be
(\Pi^{\mu\nu})_{\lambda_1}=\lambda_1 \sigma^{\langle\mu}{}_\lambda \sigma^{\nu\rangle \lambda}
\ee
where we will use the notation $(\cdots)_{\lambda_1}$ to indicate that we are only showing terms of $\cdots$ that depend on $\lambda_1$.
We will also need
\be
\sigma_{\perp z}=\sigma^{\perp z}+{\cal O}(h^2)\simeq (U'_\perp-h'_{0\perp})+\dot{h}_{\perp z}\label{sigma perp z},
\ee
and
\bea
&&\sigma_{xx}+\sigma_{yy}\simeq \tfrac 23(\tfrac 12 \dot{h}_{\perp\perp} -\dot{h}_{zz})
-\tfrac 43 (U_z'-h'_{0z})\\
&&\sigma_{xy}\simeq \dot{h}_{xy}.
\eea

The $\lambda_1$ dependence $(T^{xy})_{\lambda_1}$ of (\ref{txy}) comes solely from the $(\Pi^{xy})_{\lambda_1}$ term at the order shown:
\bea
(T^{xy})_{\lambda_1}&\simeq& (\Pi^{xy})_{\lambda_1}\simeq
\lambda_1\bigg[\sigma_{xy}(\sigma_{xx}+\sigma_{yy})+\sigma_{xz}\sigma_{yz}\bigg]
\nn\\
&\simeq& \lambda_1 \bigg[\tfrac 23 \dot h_{xy} (\tfrac 12\dot 
h_{\perp\perp}-\dot{h}_{zz}-2(U_z'-h'_{0z}))+(U_x'-h_{0x}'+\dot{h}_{xz})(U_y'-h_{0y}'+\dot{h}_{yz})\bigg].\nn\\\label{lambda1 1}
\eea
We now want to devise a simple Kubo-like formula for extracting $\lambda_1$ from some limit of a retarded 3-point correlator. To that end, note that in the limit of $\vec x$-independent sources, $h_{\mu\nu}=h_{\mu\nu}(t)$, equation (\ref{lambda1 1}) simplifies to
\be
(T^{xy})_{\lambda_1}\simeq \lambda_1 \bigg[\tfrac 23 \dot h_{xy}(\tfrac 12 \dot h_{\perp\perp}- \dot h_{zz})+\dot h_{xz}\dot h_{yz}\bigg].\label{lambda1 2}
\ee
In momentum space this becomes
\bea
(T^{xy}(\bm q))_{\lambda_1}&\simeq&
-\lambda_1 \int d^4 \bm q_1 \int d^4 \bm q_2  
 \frac{1}{(2\pi)^4}\delta^4(\bm q-\bm q_1-\bm q_2) \delta^3(\vec q_1)\delta^3(\vec q_2) \omega_1 \omega_2 \nn\\
&\times& \bigg[\tfrac 23 h_{xy}(\bm q_1)
(\tfrac 12 h_{\perp\perp}(\bm q_2)-h_{zz}(\bm q_2))+h_{xz}(\bm q_1) 
h_{yz}(\bm q_2)\bigg]\label{lambda1 22}
\eea
where the 4-momentum ${\bm  q}=(\omega, 0,0,k)$ is associated with the measurement point, and the 4-momenta $\bm q_1=(\omega_1,0,0,k_1), 
\bm q_2=(\omega_2,0,0,k_2)$ are associated with each one of the $h$ factors respectively.
 By (\ref{exp correl FT}), this produces contributions to $G^{xy|xy|\perp\perp}$,
$G^{xy|xy|zz}$ and $G^{xy|xz|yz}$ in the limit where all three spatial momenta vanish. 
For the purposes of our later AdS/CFT calculation, we will find that the least technically challenging case to calculate will be $G^{xy|xz|yz}$. So here we will focus on finding a Kubo-like formula for $\lambda_1$ in terms of $G^{xy|xz|yz}$. More Kubo-like formulas can be found in Appendix \ref{kubo}.
Consolidating identical terms in (\ref{exp correl FT}) gives
\bea
\langle T^{xy}({\bm q})\rangle_h&=&\int d^4 \bm q_1\int d^4 \bm q_2\frac{1}{(2\pi)^4}\delta^4(\bm q-\bm q_1-\bm q_2)\, 
G^{xy| xz| yz}(\bm q;-\bm q_1,-\bm q_2)h_{xz}(\bm q_1) h_{yz}(\bm q_2)\nn\\&+&
\text{other\;G's}.\label{lambda1 3}
\eea
In what follows we will suppress writing the momentum dependence of the 3-point correlator, with the understanding that the momenta $-\bm q_1, -\bm q_2$ are associated with the second pair and third pair of indices of 
the 3-point correlator respectively. 
We are now ready to compare (\ref{lambda1 3}) to (\ref{lambda1 22}) to extract $\lambda_1$:
\be
\lim_{\substack{\omega_1\to0 \\ \omega_2\to 0}}
\partial_{\omega_1}\partial_{\omega_2} \lim_{\substack{k_1\to0 \\ k_2\to 0}}G^{xy| xz| yz}=-\lambda_1+(\mbox{$\lambda_1$-independent terms}).
\ee

We leave the derivation of the $\lambda_1$-independent terms to Appendix \ref{appendix eta}. The result is:
\be
\lim_{\substack{\omega_1\to0 \\ \omega_2\to 0}}
\partial_{\omega_1}\partial_{\omega_2} \lim_{\substack{k_1\to0 \\ k_2\to 0}}G^{xy| xz| yz}=-\lambda_1+\eta\tau_\Pi.\label{lambda1 4}
\ee

\subsection{$\lambda_2$}

To evaluate the  $\lambda_2$ contribution to the stress tensor $T^{xy}$, we need to compute
\be
(T^{xy})_{\lambda_2}\simeq 
(\Pi^{xy})_{\lambda_2}
\simeq\lambda_2\tfrac 12 \delta^{xl}\delta^{ym}(\sigma_{ln}\Omega_m{}^n+\sigma_{mn}\Omega_l{}^n),\label{lambda2 0}
\ee

Assuming, as we did before, that the fluctuations depend only on $t,z$ coordinates, vorticity is given by
\begin {equation}
   \Omega_i{}^j\simeq \Omega^{ij}\simeq\tfrac 12(\partial_i U_j - \partial_j U_i)
   = \tfrac 12\begin {pmatrix}
        0 & 0 & -U'_x \\
        0 & 0 & -U'_y \\
        U'_x & U'_y & 0
     \end {pmatrix} \label{Omega}.
\end {equation}

Substituting (\ref{Omega}) and (\ref{sigma perp z}) into (\ref{lambda2 0}) 
we find
\be
(T^{xy})_{\lambda_2}\simeq 
\tfrac 14 \lambda_2 \bigg(-U'_x(U'_y-h'_{0y}+\dot{h}_{yz})-U'_y(U'_x-h'_{0x}+\dot{h}_{xz})\bigg).\label{lambda2 2}
\ee
Again, we are interested in finding a simple Kubo relation which is amenable to a straightforward AdS/CFT computation. We settle on using
$G^{xy|yz|0x}$ and we consider sources such that $h_{yz}=h_{yz}(t)$ and $h_{0x}=h_{0x}(z)$. After a substitution of $U_\perp$ from (\ref{eq:uperp}) into (\ref{lambda2 2}), and a comparison with the relevant terms from (\ref{exp correl FT})
\bea
\langle T^{xy}(\bm q)\rangle_h&=&\int d^4 \bm q_1\int d^4 \bm q_2 \frac{1}{(2\pi)^4}\delta^4(\bm q-\bm q_1-\bm q_2)\,  
\>G^{xy| yz| 0x}(\bm q;-\bm q_1,-\bm q_2)h_{yz}(\bm q_1) 
h_{0x}(\bm q_2)\nn\\&+&\mbox{other $G$'s}
\eea
we are led to
\be
\lim_{\substack{\omega_1\to 0\\k_2\to 0}}\partial_{k_2}\partial_{\omega_1}\lim_{\substack{\omega_2\to 0\\k_1\to 0}}G^{xy| yz| 0x}=-\tfrac 14\lambda_2+\tfrac 12 \eta\tau_{\Pi},\label{lambda2 3}
\ee
where the $\lambda_2$-independent term $\tfrac 12\eta \tau_\Pi$ is derived in
Appendix \ref{appendix eta}.
Other Kubo-type relations can be found in Appendix \ref{kubo}.

\subsection{$\lambda_3$}
$\lambda_3$ is the coefficient of the square of the vorticity tensor in the hydrodynamic expansion of the stress tensor. To get $\lambda_3$, we rely on computing the same component $T^{xy}$:
\begin {equation}
   (T^{xy})_{\lambda_3} \simeq 
   (\Pi^{xy})_{\lambda_3}.
\simeq \tfrac 14 \lambda_3 U'_x U'_y
\end {equation}
Substituting (\ref{eq:uperp}) we get
\begin {equation}
   (T^{xy})_{\lambda_3}
   \simeq
   -\tfrac{1}{4} \lambda_3 k_1 k_2
      \frac{D_\eta k_1(k_1 h_{0x} + \omega_1 h_{zx})}{(-i \omega_1 + D_\eta k_1^2)}
      \frac{D_\eta k_2(k_2 h_{0y} + \omega_2 h_{zy})}{(-i \omega_2 + D_\eta k_2^2)}.\label{lambda3 1}
\end {equation}
A simple way to extract $\lambda_3$  from terms up to second order in momenta in the 3-point correlator (at higher order in momenta we would need to expand the stress tensor to terms that include higher order hydro coefficients) is to take the static limit: $\omega_1=\omega_2=0$.
Then, a comparison of (\ref{lambda3 1}) with the stress tensor response  in terms of Green's functions (\ref{exp correl FT}), namely
\begin {equation}
  \langle T^{xy}(\bm q) \rangle_h
  = \int d^4\bm q_1
\int d^4\bm q_2 \frac{1}{(2\pi)^4}\delta^4(\bm q-\bm q_1-\bm q_2) 
      G^{xy| 0x| 0y}(\bm q;-\bm q_1,-\bm q_2) h_{0x}(\bm q_1)   h_{0y}(\bm q_2)
    + \mbox{other $G$'s} ,
\end {equation}
yields
\begin {equation}
     \lim_{\substack{k_1\to0\\k_2\to0}}
     \partial_{k_1} \partial_{k_2}
      \lim_{\substack{\omega_1\to0\\\omega_2\to0}}
            G^{xy| 0x| 0y}
   =
   -\tfrac14 \lambda_3 . \label{lambda3 2}
\end {equation}
We will use AdS/CFT to evaluate $\lambda_3$ from (\ref{lambda3 2}). However, for the reader's convenience we give other Kubo formulae in Appendix \ref{kubo}.
\section{Stress tensor correlators via AdS/CFT}
The generating functional for the stress tensor correlators of ${\cal N}{=}4$ super Yang-Mills theory in the limit of large number of colors ($N_c\gg1$) and at strong coupling is the on-shell five-dimensional gravitational action composed of  the Einstein-Hilbert action, a cosmological constant term, the Gibbons-Hawking term, and holographic renormalization counterterms
 \cite{Balasubramanian:1999re, Kraus:1999di, Emparan:1999pm, de Haro:2000xn}:
\be
{\cal S}=\frac{N_c^2}{8\pi^2 R_{\rm AdS}^3}\bigg[\int_{\cal M}\sqrt{-g}( R-2\Lambda)+2\int_{\partial {\cal M}}\sqrt{-g_{\text{bdy}}} K+a\int_{\partial {\cal M}}\sqrt{-g_{\text{bdy}}}-\frac{R_{\rm AdS}}{d-2}\int_{\partial {\cal M}}\sqrt{-g_{\text{bdy}}} R_{\text{bdy}}\label{act2}\bigg]
\ee
where the values of the cosmological constant and of the volume counterterm parameter $a$ are
\be
\Lambda=-\frac{d(d-1)}{2R_{\rm AdS}^2}\;\;\text{and}\;\; a=-\frac{2(d-1)}{R_{\rm AdS}} \label{lambda a}, \;\;\text{with} \;\;d=4.
\ee
The trace of the extrinsic curvature tensor can be expressed in terms of the induced metric on the boundary $(g_{\text{bdy}})_{MN}=g_{MN}-n_M n_N$ and the unit normal to the boundary $n^M$ as 
\be
K=(g_{MN}-n_M n_N)\nabla^M n^N.
\ee
In general, ${\cal M}$ asymptotes to an AdS space of radius $R_{\rm AdS}$. Since we are studying finite-temperature super Yang-Mills theory, the background ${\cal M}$ is five-dimensional AdS-Schwarzschild space \cite{Witten:1998zw}:
\be
ds^2=\frac{(\pi TR_{\rm AdS})^2}{u}(-f(u) dt^2 +\vec x^2)+\frac{R_{\rm AdS}^2 du^2}{4u^2 f(u)}, \qquad f(u)=1-u^2.
\ee
Taking advantage of the fact that the AdS radius drops out of final results, it is convenient to set $R_{\rm AdS}=2$ and work in units $2\pi T=1$ so that 
\be
ds^2= \frac{-f(u) dt^2+d\vec x^2}{u}+\frac{du^2}{u^2 f(u)}\label{bh metric}.
\ee 

Next we will evaluate the on-shell action, by expanding the metric in fluctuations $\delta g_{MN}$ (where $M, N=0,1,2,3,5=t,x,y,z,u$) around the AdS-Schwarzschild background. 
In imaginary time, AdS-Schwarzschild is smooth and non-singular, and 3-point correlators are given by Witten diagrams as depicted in Fig.~1a.
In real time, as discussed in \cite{Arnold:2010ir,Barnes:2010jp}, the corresponding diagrams for \underline{retarded} correlators $G_{\rm raa}$ live in the right-quadrant of the Penrose diagram due to causality. The lines in this diagram represent advanced and retarded bulk-to-boundary propagators. These propagators are identified as solutions $\delta g_M^N$ to the linearized equations of motion, which approach prescribed values at the boundary $u=0$ and which reduce (up to gauge terms) to purely incoming/outgoing waves: $e^{-i\omega t} (1-u)^{\pm i\omega/2}$.

\subsection{Two-point stress tensor correlators and comparison with hydrodynamic expansion}
First we will warm up with 2-point retarded correlators. 
We recall that the gravity fluctuations are taken to be independent of $x,y$ coordinates, and are slowly varying functions of $t, z$. Given the symmetry of the problem, the fluctuations can be classified according to their transformations under rotations about the $z$ axis as: $SO(2)$ tensors ($\delta g_{xy}$ and $\delta g_{xx}-\delta g_{yy}$), vectors ($\delta g_{0x}, \delta g_{zx}$ and
$\delta g_{0y}, \delta g_{zy}$), and scalars (all others).

For completeness we list in Appendix \ref{Bb prop} the bulk metric fluctuations in momentum space as a series expansion in $\omega, k$, and in terms of the boundary metric fluctuations, $h_\mu^\nu$. The  retarded  bulk-to-boundary graviton propagators are easily obtained from the expressions given in Appendix \ref{Bb prop} by differentiating with respect to the boundary fields. We work in the gauge
\be
\delta g_{M5}=0.
\ee

The tensor modes $\delta g_x^y$ and $\delta g_x^x -\delta g^y_y$ have propagators which do not have singularities for $\omega, k\ll 1$ \cite{Policastro:2002se}. The vector (or shear) modes, e.g. $\delta g^x_0, \delta g^x_z$, have poles typical of diffusion, and the scalar mode propagators have a sound pole \cite{Policastro:2002tn}. We will refer to the vector metric fluctuations as shear modes, and to the scalar metric fluctuations as sound modes. 

The computation of the stress tensor 1-point function is reviewed in Appendix \ref{1pt}.1\footnote{Restoring the units, the energy density and pressure equal $\bar \epsilon=\frac{3\pi^2 N_c^2 T^4}{8}, \bar P=\frac{\pi^2 N_c^2 T^4}{8}.$}:
\be
\langle T^{00}\rangle_0 =2\frac{\delta {\cal S}}{\delta h_{00}}= \frac{3 N_c^2}{2^7 \pi^2}\equiv \bar \epsilon \label{bar epsilon}
\ee
gives the energy density of the finite temperature strongly coupled super Yang-Mills theory and 
\be
\langle T^{ij}\rangle_0 = 2\frac{\delta {\cal S}}{\delta h_{ij}}=\delta^{ij} \frac{N_c^2}{2^7 \pi^2}\equiv \delta^{ij}\bar P , \qquad i,j=1,2,3=x,y,z,
\ee
gives its pressure.

The two-point retarded correlators are computed using the quadratic gravitational vertex reviewed in Appendix E.2. Since we are interested in $\rm {ra}$ correlators, one of the bulk-to-boundary propagators is retarded (the one associated with the largest time point on the boundary), and the other propagator is advanced. However, we should remember that the bulk-to-boundary propagators in momentum space are defined as retarded or advanced relative to the momentum conjugate to the boundary space-time point. 4-momentum conservation gives $\bm q = -\bm q_1$, where $\bm q$ is the momentum associated with the point which has the largest time. This effectively transforms the advanced bulk-to-boundary propagator ${\cal G}_{\rm ar}(\bm q_1, u)$ into a retarded propagator ${\cal G}_{\rm ra}(\bm q, u)$.

From the decoupled fluctuation $\delta g^x_{y}$ one recovers\footnote{For the sake of brevity, we suppress energy-momentum conservation delta functions.} the stress tensor retarded 2-point function \cite{Baier:2007ix}
\bea
G_{\rm AdS}^{xy| xy}&=&-\frac{\delta^2 {\cal S}}{\delta^2 h_{xy}}\nn\\
&=&\frac{N_c ^2}{2^7\pi^2}-i\frac{N_c^2 \omega}{2^6\pi^2}
+\frac{(\omega^2(1-\ln 2)-k^2)N_c^2}{2^6\pi^2}+\cdots\label{xyxy ads}
\eea
The result derived in the hydrodynamic limit from solving 
(\ref{cons law}) and (\ref{conf}) is 
\be
G_{\text {hydro}}^{xy| xy}=\tfrac 13 \bar\epsilon
-i\eta\omega +\eta\tau_\Pi\omega^2-\tfrac 12\kappa(\omega^2+k^2)+\cdots\label{xyxy hydro}
\ee
where dots denote terms which are higher order in gradients (and hydrodynamic expansion coefficients). By identifying (\ref{xyxy ads}) and (\ref{xyxy hydro})
one obtains a handful of hydro coefficients, of first and second order\footnote{Restoring the units, this reads  $\eta = \frac{\pi N_c^2 T^3}{8}, 
\kappa = \frac{N_c^2 T^2}8, 
 \eta\tau_\Pi = \frac{N_c^2 (2-\ln 2) T^2}{16}.$}:
\be
\eta = \frac{N_c^2}{2^6\pi^2}, 
\qquad \kappa = \frac{N_c^2}{2^5\pi^2}, 
\qquad \eta\tau_\Pi = \frac{N_c^2 (2-\ln 2)}{2^6 \pi^2}. \label{hydro 2nd}
\ee
From (\ref{bar epsilon}) and (\ref{hydro 2nd}), the diffusion constant of the shear modes in the strongly coupled ${\cal N}{=}4$ super Yang-Mills plasma is
\be
D_{\eta}=\frac{\eta}{\bar\epsilon+\bar P}=\frac 12.
\ee
This is the maximal set of hydrodynamic coefficients which can be determined from 2-point stress tensor correlators in the AdS-Schwarzschild background \cite{Baier:2007ix}. On the other hand, having found
$\bar\epsilon,\bar P,\eta,\tau_\Pi$ and $\kappa$, we should be able to match all the other 2-point correlators obtained via AdS/CFT with their hydrodynamic expressions.

To see how this plays out we consider the 2-point functions of shear modes. We perform an expansion assuming a shear dispersion relation $\omega\sim k^2\sim \lambda^2\ll1$, where $\lambda$ is a small expansion parameter, and extend the results of Policastro, Son and Starinets \cite{Policastro:2002se} to second order in $\lambda$:
\bea
G_{\rm AdS}^{0x| 0x}&=&-\frac{\delta^2{\cal S}}{\delta h_{0x} \delta h_{0x}}
\nn\\&=&
\frac{N_c^2}{2^4\pi^2}\bigg[-\frac{k^2+6i \omega}{2^4(-i\omega+\tfrac 12 k^2)}+
\frac{k^2(2\omega^2(\ln2-1)-2i\omega k^2+k^4)}{2^3(-i\omega+\tfrac 12 k^2)^2}+\cdots\bigg]\nn
\\
G_{\rm AdS}^{0x| xz}&=&-\frac{\delta^2{\cal S}}{\delta h_{0x}\delta h_{xz}}\nn\\
&=&
\frac{N_c^2}{2^4 \pi^2}\bigg[-\frac{\omega k}{2^2(-i\omega+\tfrac 12 k^2)}+\cdots\bigg]\nn
\\
G_{\rm AdS}^{xz| xz}&=&-\frac{\delta^2{\cal S}}{\delta h_{xz} \delta h_{xz}}\nn\\
&=&
\frac{N_c^2}{2^4 \pi^2}\bigg[\frac 1{2^3}-\frac{\omega^2}{2^2(-i\omega+\tfrac 12 k^2)}+\cdots\bigg].
\eea 
As a result of this expansion, the shear poles are visible.
The presence of the higher order poles in the 2-point correlators is an artifact of the expansion and indicates that the location $\omega=-i\tfrac 12 k^2$ of the shear pole is shifted by higher order terms\footnote{The shear pole, as extracted from second order hydro, is \cite{Baier:2007ix}: $\omega\simeq -i D_\eta k^2-D_\eta\tau_\Pi \omega k^2.$}.
 
This can be contrasted with
\bea
G_{\text {hydro}}^{0x| 0x}&=&-\frac{\bar \epsilon(i\omega+\tfrac 13 D_\eta k^2)}{-i\omega+D_\eta k^2}+\frac{\omega k^2 (-(\eta \tau_\Pi-\tfrac 12\kappa)\omega+\tfrac 12 i \kappa D_\eta k^2)}{(-i\omega+D_\eta k^2)^2}+\cdots
\nn\\
G_{\text {hydro}}^{0x| xz}&=&-\frac{\eta\omega k}{-i\omega+D_\eta k^2}+\dots
\nn\\
G_{\text {hydro}}^{xz| xz}&=&\tfrac 13
\bar\epsilon-\frac{\eta\omega^2}{-i\omega+D_\eta k^2}+\cdots
\eea
There is an apparent mismatch in the hydrodynamic expansion and the AdS calculation in $G^{0x| 0x}$, namely the coefficients $\omega k^4$ and $k^6$ which multiply the second order shear pole disagree. This puzzle is resolved by noting that the expansion in small $\omega, k$ assuming a shear dispersion relation has the rather unwanted effect of mixing  hydrodynamic coefficients of different order in each term of the series expansion: e.g. $\bar\epsilon$ and $\eta$ at zeroth order in $\lambda$; $\eta\tau_\Pi-\tfrac 12 \kappa$ and $D_\eta \kappa$ at second order in $\lambda$; etc. So a contamination with third-order hydro coeffcients is to be expected, as forecast by the term proportional with 
$D_\eta\kappa$ which contains the product of two second order hydro coefficients. This is further elucidated by a comparison with what happens when instead expanding the correlators assuming $\omega\sim k\sim\lambda\ll1$. Then, the AdS result is 
\bea
G_{\rm AdS}^{0x| 0x}&=&\frac{N_c^2}{2^4\pi^2}\bigg[\frac{3}{2^3}
-i\frac{ k^2 }{2^2\omega}
+\frac{k^2 (2\omega^2(1-\ln2)-k^2)}{2^3\omega^2}+\cdots\bigg]\nn
\\
G_{\rm AdS}^{0x| xz}&=&\frac{N_c^2}{2^4\pi^2}\bigg[-i\frac{k }{2^2}
+\frac{k(2\omega^2(1-\ln 2)-k^2)}{2^3\omega}+\cdots\bigg]\nn
\\
G_{\rm AdS}^{xz| xz}&=&\frac{N_c^2}{2^4\pi^2}\bigg[\frac{1}{2^3}-i\frac{\omega}{2^2}
+\frac{(2\omega^2(1-\ln 2)-k^2)\pi^2 N_c^2 T^4}{2^3}+\cdots\bigg]
\eea
while the hydro expansion gives
\bea
G^{0x| 0x}_{\text{hydro}}&=&\bar\epsilon -i\frac{\eta k^2}{\omega}+
\frac{k^2(\eta\tau_\Pi\omega^2-\tfrac 12\kappa \omega^2-\eta D_\eta k^2)}{\omega^2}+\cdots\nn\\
G^{0x| xz}_{\text{hydro}}&=&-i\eta k + \frac{k(\eta\tau_\Pi \omega^2-\tfrac 12 \kappa \omega^2
-\eta D_\eta k^2)}{\omega}+\dots\nn\\
G^{xz| xz}_{\text{hydro}}&=&\tfrac 13\bar\epsilon -i\eta\omega 
+\eta\tau_\Pi\omega^2-\tfrac 12 \kappa\omega^2-\eta D_\eta k^2+\cdots
\eea
The $\omega$ poles are a remnant of expanding the shear pole, but now the two sets of 2-point correlators are perfectly matched, and at each term in the expansion there is no mixing between different order hydro coefficients. 
One may extract  $\bar\epsilon,\eta$ and the combination $\eta\tau_\pi-\tfrac 12\kappa$ with the same result as before.
In this respect, the tensor mode 2-point correlator $G^{xy| xy}$ enables the identification of a larger set, since $\eta\tau_\Pi$ and $\kappa$ can be obtained separately.

In appendix \ref{sound} we list the 2-point functions of sound modes, computed using a sound mode dispersion expansion $\omega\sim k\sim\lambda\ll1$ to second order in $\lambda$, extending the results of Policastro, Son and Starinets \cite{Policastro:2002tn}. The higher-order poles which appear in appendix \ref{sound}  
are also an artifact of expanding the higher-order attenuation terms which are present in the sound pole. There is no mixing of the different hydro coefficients at each term in the expansion, and the hydro and AdS results match, provided that one takes into account that the AdS computation yields tensor density correlators ($2(\delta^2{\cal S})/(\delta h_{\mu\nu}\delta h_{\rho\sigma})=\delta \langle \sqrt{-g} T_{\mu\nu}\rangle/\delta h_{\rho\sigma}$) and not tensor correlators ($\delta \langle T^{\mu\nu}\rangle/\delta h_{\rho\sigma}$).   
The difference is a contact term, namely
\be
G_{\text{hydro}}^{\mu\nu,\rho\sigma}=G_{\rm AdS}^{\mu\nu,\rho\sigma}+\eta^{\rho\sigma} T^{\mu\nu},\label{diff cor}
\ee
which arises from differentiating the volume factor $\sqrt{-g}$ with respect to the metric fluctuations.

\subsection{$\lambda_1,\lambda_2$ and $\lambda_3$ via AdS/CFT}

Due to the complexity of evaluating generic stress tensor 3-point functions, in this section we contend ourselves with computing $\lim_{\substack{k_1\to 0\\k_2\to0}}G^{xy| yz| xz}_{\rm AdS}$, $\lim_{\substack{\omega_1\to 0\\\omega_2\to0}}G^{xy| 0y| 0x}_{\rm AdS}$ and $\lim_{\substack{k_1\to 0\\\omega_2\to0}}G^{xy| yz| 0x}_{\rm AdS}$. We leave a more complete computation of 3-point stress tensor correlators at arbitrary (albeit small) values of 4-momenta for future work.

Note that we are using 3-point correlators where one pair of legs, $xy$ (which is associated with the largest-time boundary point) corresponds to tensor fluctuations,  and the other leg pairs correspond to $O(2)$ vector (shear mode) fluctuations: $xz,x0$ and $yz,y0$. 
Because of the mixing between $\delta g_{\perp 0}$ and $\delta g_{\perp z}$, each of the corresponding bulk-to-boundary propagators is a $2\times2$ matrix. So, for example, the Witten diagram for $G^{xy|yz|xz}$ has four terms, depicted schematically in Fig. 2. 

\begin{figure}
\begin{center}
\includegraphics[width=3.6in]{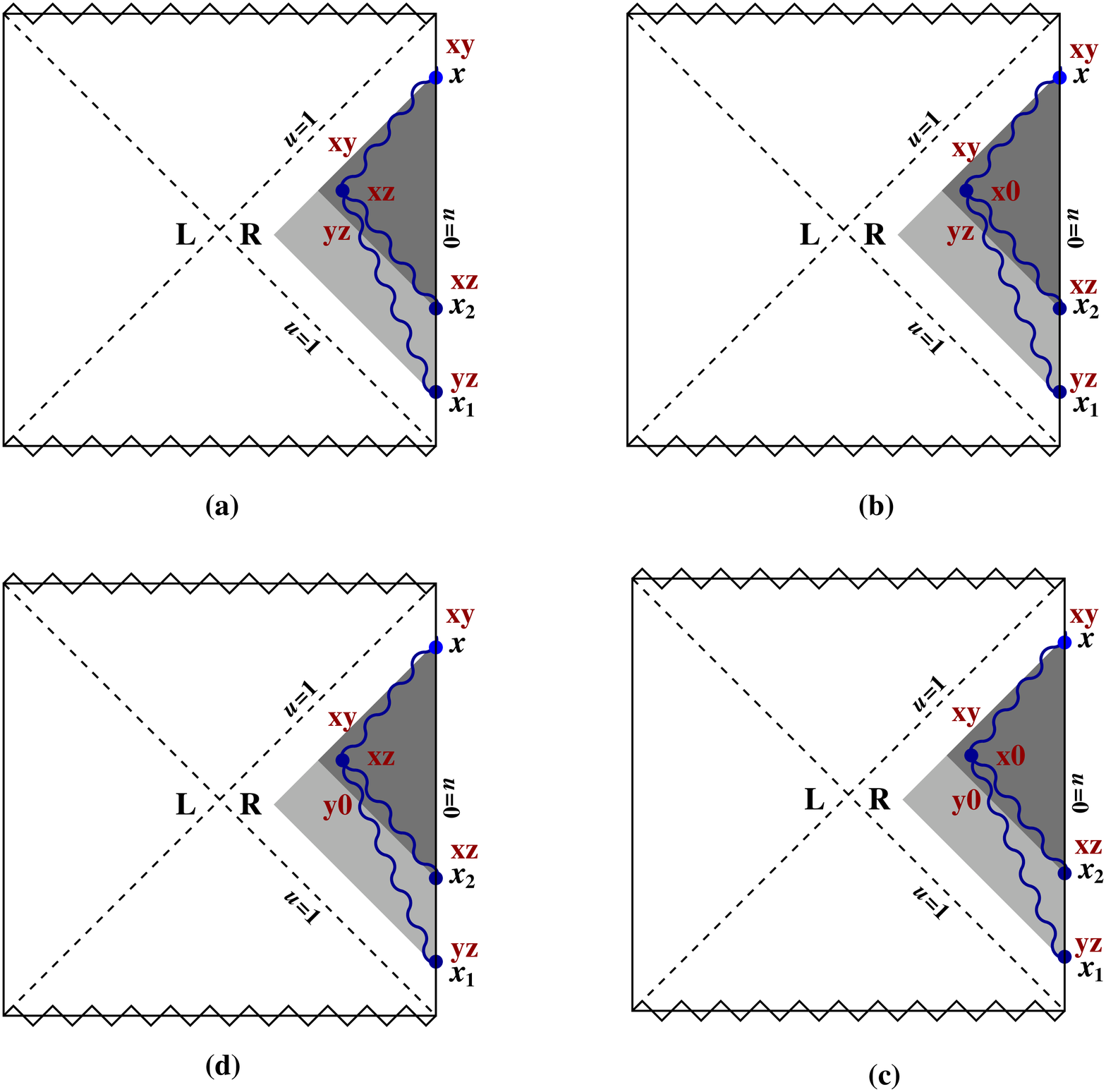} \\
Figure 2: Witten diagrams for the retarded 3-point correlator $G^{xy|yz|xz}$ with the boundary point $\bm x$ having the largest time; $\bm x_1$ and $\bm x_2$ can have any time order.
\end{center}
\end{figure}

Moreover, each $2\times2$ shear-mode propagator contains a pure gauge (diffeomorphism) contribution whose behavior at the horizon is not in the form of an incoming/outgoing wave. This gauge artifact can (and does) cause technical difficulties in integrating the location of the vertex in the Witten diagram up to the horizon\footnote{The problematic behavior is related to the fact that the diffeomorphism parameters (as constrained by the gauge condition $\delta g_{M 5}=0$) are non-analytic functions at the horizon (see equation (\ref{xi})).}. We will avoid this issue by computing correlators for which the diffeomorphism terms drop out, that is, we will be focus on calculations where we can easily work with gauge-invariant modes.

As an example, the low momentum solution to the equation of motion for $\delta h_0^x$ and $\delta g_z^x$ which gives the retarded bulk-to-boundary propagators
is\footnote{Here, since we are interested in sending the spatial momentum to 0 first, we give the shear mode propagators
assuming a sound mode dispersion relation. In Appendix \ref{Bb prop} we assumed a diffusion dispersion relation to highlight the presence of the diffusion pole.} 
\bea
\delta g^x_0&=&\tilde C_3 (1-u)^{-i\omega/2}\bigg(i\frac{kf}{2}+\omega k(-\frac f4 \ln(1+u)+\frac{\ln 2}2 -\frac u2 +\frac {(1-\ln2)u^2}2)+\cdots\bigg)
+\tilde D_1\nn\\
\delta g^x_z&=&\tilde C_3 (1-u)^{-i\omega/2}\bigg(1-i\frac{\omega}2\ln(1+u)+\omega^2 (-\frac 12 \text{Li}_2(\frac{1-u}2) + \frac 18 \ln^2(1+u)\nn\\
&+&(1-\frac {\ln 2}2)\ln(1+u))+\cdots\bigg)
-\tilde D_1\frac{k}\omega
\eea
where
\bea
\tilde C_3&=&\bigg(\frac k\omega-i\frac{k^3}{2\omega^2}-\frac{k(\omega^4(6 \ln^2{2} -\pi^2)+12\omega^2 k^2\ln2+6k^4)}{24\omega^3}+\cdots\bigg)(h^x_{0}+\frac{\omega}{k} h^x_{z})\nn\\
\tilde D_1&=&\bigg(1-i\frac{k^2}{2\omega}-\frac{k^2(2\omega^2\ln2+k^2)}{4\omega^2}+\dots\bigg)h^x_0\nn\\
&+&\bigg(-i\frac{k}2-\frac{k (2\omega^2\ln2+k^2)}{4\omega}+\cdots\bigg)h^x_3 .
\label{tilde d1}
\eea
The $\tilde D_1$ terms in (\ref{tilde d1}) represent a gauge mode associated with shifts in the $x$ direction. One possible way to trivially isolate the gauge mode is to take the $k\to 0$ limit. Then the the shear modes decouple. $\delta g^x_0(k{=}0)$ is equal to the boundary value up to $O(\omega^3)$ corrections,  and $\delta g^x_z(k{=}0)$ is a purely incoming wave at the horizon to the same $O(\omega^3)$ order:
\bea
\delta g^x_0(k{=}0)&=& h^x_0(k{=}0)+\dots,
\nn\\\delta g^x_z(k{=}0) &=& h^x_z(k{=}0)\times\bigg(1-\frac{\omega^2(6\ln^2 2-\pi^2)}{24}+\dots\bigg)(1-u)^{-i\omega/2}\bigg(1-i\frac{\omega}2\ln(1+u)\nn\\
&+&\omega^2 (-\frac 12 \text{Li}_2(\frac{1-u}2) + \frac 18 \ln^2(1+u)
+(1-\frac {\ln 2}2)\ln(1+u))+\cdots\bigg).\label{causal omega}
\eea
As a result, only Fig.~2a contributes to $G_{\rm AdS}^{xy| xz| yz}(k_1{=}0,k_2{=}0)$ [a consequence of $O(3)$ rotation invariance when $k=0$], and we expect no subtleties coming from the near horizon region.

Except for the specific form of the cubic gravitational vertex, the evaluation of the 3-point stress tensor correlator is no different than evaluating a causal scalar 3-point correlator with a derivative cubic vertex in the AdS-Schwarzschild bulk\footnote{If the scalar supergravity fields have a derivative coupling of the type $\int_{\cal M} \sqrt{-g} g^{MN}\partial_M \phi \partial_N \phi$, then, for on-shell fields this reduces to  $(m^2/2) \int_{\cal M} \sqrt{-g}\phi^3$ plus a total derivative term. However, in the text, we refrain from performing the integration by parts and merely comment on the behavior of the integrand near the horizon. }. 
Consider this scalar example, with cubic interaction $\int_{\cal M} \sqrt{-g} g^{MN}\partial_M \phi \partial_N \phi$, for the sake of simplicity of discussion, and examine the small frequency (with $k$'s zero) behavior of this contribution to the action ${\cal S}$. As we'll see, this is dominated by the near-horizon ($u\to1$) contribution
\be
\sqrt{-g} g^{uu} \phi \partial_u \phi \partial_u \phi.\label{interm phi}
\ee
The small frequency behavior of the retarded scalar bulk-to-boundary propagator is $(1-u)^{-i\omega/2}$.
Therefore, computing the small frequency behavior of the integrand in (\ref{interm phi}) gives a contribution to the causal retarded correlator which in the near-horizon region is proportional to 
\be
\int du 
(1-u)^{-i(\omega_1+\omega_2)-1} (\omega_1\omega_2 - (\omega_1+\omega_2)^2)\label{interm phi1}
\ee
where the terms quadratic in $\omega$'s arise from taking the $u$ derivatives.
In arriving at (\ref{interm phi1}) we took the momenta flowing through the $\rm a$ legs of the $\rm {raa}$ correlator to be $\omega_1$ and $\omega_2$. By energy conservation, the momentum flowing through the $\rm r$ leg is $\omega=-(\omega_1+\omega_2)$. The potential divergence at the horizon is regularized by the fact that the causal 3-point function of interest corresponds to $\omega_1\to \omega_1 - i\epsilon, \omega_2\to \omega_2-i\epsilon$ and $\omega=-(\omega_1+\omega_2)\to -(\omega_1+\omega_2)+2i\epsilon$.\footnote{ See the discussion in sec II.B of \cite{Arnold:2010ir}, and see \cite{Evans:1990qh,Evans:1991ky}.}
 Integrating (\ref{interm phi1}) from $u=0$ to 1 yields
\be
\frac{i}{\omega_1+\omega_2}(\omega_1\omega_2 - (\omega_1+\omega_2)^2)
\ee
to leading order in $\omega$'s. Counting powers of frequency as
$\lambda\sim\omega_1\sim\omega_2$, the near-horizon divergence of the integrand has enhanced the naively $O(\lambda^2)$ order of the integrand to an $O(\lambda)$ contribution to the 3-point function.

A similar story emerges for the $G_{\rm AdS}^{xy| xz| yz}(k_1=0,k_2=0)$ correlator. There, the integrand, as given by the expansion of the cubic gravitational action, behaves near the horizon as
\be
\frac{3N_c^2}{2^6}(1-u)^{-i(\omega_1+\omega_2)}\bigg(\frac{(\omega_1+\omega_2)^2}{
(1-u)}-i\frac{\ln 2(\omega_1+\omega_2)^3}{(1-u)}+\cdots\bigg)\label{interm 3pt}
\ee 
where at least one of the $\omega$ factors arise from derivatives on the wave factor $(1-u)^{-i(\omega_1+\omega_2)}$.\footnote{As a result, finding the 
integrand in (\ref{interm 3pt}) only requires knowing the solution $\delta g^{(1)}$ to the linearized equations of motion to $O(\omega_1^2,\omega_2^2)$, as given in (\ref{interm phi1}).} 
Regularizing the wave factors as before, and performing the integral over the black hole bulk up to the horizon, gives a finite contribution to the $G_{\rm AdS}^{xy| xz| yz}(k_1{=}0,k_2{=}0)$ correlator:
\be
\lim_{\substack{k_1\to 0\\k_2\to0}}G^{xy| yz| xz}_{\rm AdS}=\frac{N_c^2}{2^4 \pi^2}\bigg[
\frac 1{2^3}-i\frac{\omega_1+\omega_2}{2^2}-\frac{(\omega_1\omega_2+\omega_1^2+\omega_2^2)(\ln2-1)}{2^2}+\cdots\bigg].
\ee
The last step is to identify the AdS/CFT correlator 
with the corresponding hydro expansion\footnote{ Given the discussion preceding
(\ref{diff cor}),
readers may wonder why we are identifying tensor density correlators ($G_{\rm{AdS}}$) with tensor 
correlators ($G_{\rm{hydro}}$). There is no difference for the particular components we are interested. In general,
\bea
&&G^{\mu\nu|\rho\sigma|\tau\zeta}_{\rm hydro}(\bm x,\bm y,\bm z)=
G^{\mu\nu|\rho\sigma|\tau\zeta}_{\rm AdS}(\bm x,\bm y,\bm z)+
G^{\mu\nu|\rho\sigma}_{\rm AdS} (\bm x,\bm y)g^{\tau\zeta}\delta(\bm x-\bm z)
+G^{\mu\nu|\tau\zeta}_{\rm AdS} (\bm x,\bm z)g^{\rho\sigma}\delta(\bm x-\bm y)\nn\\
&&+G^{\mu\nu}_{\rm AdS}(\bm x)(g^{\rho\sigma}g^{\tau\zeta}+g^{\rho\tau}g^{\sigma\zeta}
+g^{\rho\zeta}g^{\sigma\tau})\delta(\bm x-\bm y)\delta(\bm x-\bm z).\nn
\eea}
\be
\lim_{\substack{k_1\to 0\\k_2\to0}}G^{xy| yz| xz}_{\text {hydro}}=\tfrac13\bar \epsilon
-i\eta(\omega_1+\omega_2)+ \eta\tau_\Pi(\omega_1^2+\omega_2^2+\omega_1\omega_2)-\tfrac12 \kappa(\omega_1^2+\omega_2^2)-\lambda_1\omega_1\omega_2+\dots
\ee
to get $\lambda_1$:
\be
\lambda_1=\frac{N_c^2}{2^6\pi^2}.
\ee

For the remaining $\lambda_2,\lambda_3$ coefficients we will need to take the limit $\omega\to0$ of the shear modes. From Appendix \ref{Bb prop} we get that
$\delta g^x_z(\omega{=}0)$ is equal to its boundary value to order $O(k^3)$, and $\delta g^x_0(\omega{=}0)$ decouples:
\bea
\delta g^x_0(\omega{=}0)&=&\bigg(1-u^2-k^2 u(1-u)+\dots\bigg) h^x_0(\omega{=}0)\nn\\
\delta g^x_z(\omega{=}0)&=&h^x_z(\omega{=}0)+\cdots
\eea
Thus, once again, the graviton shear modes decouple, and the correlators needed to evaluate $\lambda_2$ and $\lambda_3$ are given  by a single Witten-type diagram, with the causal graviton propagators being diagonal in both the tensor and shear mode sectors.
The integral over the cubic action is straightforward and gives:
\be
\lim_{\substack{\omega_1\to 0\\\omega_2\to0}}G^{xy| 0y| 0x}_{\rm AdS}=\frac{N_c^2}{2^6 \pi^2}\bigg[-\frac1{2}+(k_1^2+k_2^2)+\cdots\bigg].
\ee
Comparison with the hydro expansion
\be
\lim_{\substack{\omega_1\to 0\\\omega_2\to0}}G^{xy| 0y| 0x}_{\text {hydro}}=-\tfrac 13\bar\epsilon+\tfrac 12\kappa (k_2^2+k_1^2)-\tfrac 14 \lambda_3k_1 k_2+\cdots
\ee
yields
\be
\lambda_3=0.
\ee
Similarly, we compute
\be
\lim_{\substack{k_1\to 0\\\omega_2\to0}}G^{xy| yz| 0x}_{\rm AdS}=\frac{N_c^2}{2^6\pi^2}
\omega_1 k_2+\cdots
\ee
which can be identified with 
\be
\lim_{\substack{k_1\to 0\\\omega_2\to0}}G^{xy| yz| 0x}_{\text {hydro}}= (-\tfrac 14\lambda_2+\tfrac 12\eta\tau_{\Pi})\omega_1 k_2+\cdots
\ee
to yield\footnote{See \cite{Haack:2008xx} for arguments regarding a universal 
relationship between $\lambda_1$, $\lambda_2$ and $\eta\tau_\Pi$.},
\be
\lambda_2 = -\frac{N_c^2}{2^5 \pi^2}\ln2.
\ee
This concludes our derivation of the second order hydrodynamic coefficients $\lambda_1$, $\lambda_2$ and $\lambda_3$ from 3-point stress tensor correlators\footnote{In addition to the computation detailed in the main text, we have performed a separate check of (\ref{preview2}) using a different set of gauge-invariant modes e.g. choosing the boundary fields such that $h_x^x=-h_y^y$ or 
$h_0^0=-\tfrac 13 k^2 (h_x^x+h_y^y)+\cdots, h_z^z=(-1+k^2)(h_x^x+h_y^y)+\cdots, h_0^z=-\tfrac 12 (ik + k\omega \ln 2)(h_x^x+h_y^y)+\cdots$ where ellipsis denotes that we have required that the gauge component of the bulk-to-boundary propagators vanishes to order $O(\lambda^3)$ in a small $\omega\sim k\sim\lambda \ll1$ expansion. 
For example, solving the hydrodynamic equations we compute 

$G^{xx-yy|0x|0x}_{\rm hydro}(\omega_1{=}\omega_2{=}0)=-\tfrac 23 \bar\epsilon +\kappa (k_1^2+k_2^2) - \tfrac 12 \lambda_3 k_1 k_2+\cdots,$

$G^{xx+yy-2zz|xy|xy}_{\rm hydro}(k_1{=}k_2{=}0)=\tfrac 43 \bar\epsilon-4i \eta (\omega_1+\omega_2)+4\eta\tau_\Pi (\omega_1^2+\omega_2^2+\omega_1\omega_2)-2\kappa^2(\omega_1^2+\omega_2^2)-4\lambda_1\omega_1\omega_2+\cdots,$

$G^{xx-yy|zx|0x}_{\rm hydro}(k_1{=}0,\omega_2{=}0)=(-\tfrac 12 \lambda_2 + \eta\tau_\Pi)\omega_1 k_2+\cdots.$

\noindent The corresponding AdS 3-point functions are 

$G^{xx-yy|0x|0x}_{\rm AdS}(\omega_1{=}\omega_2{=}0)=-\frac{N_c^2}{2^6\pi^2}+
\frac{N_c^2}{2^5\pi^2}(k_1^2+k_2^2)+\cdots,$

$G^{xx+yy-2zz|xy|xy}_{\rm AdS}(k_1{=}k_2{=}0)=\frac{N_c^2}{2^5\pi^2}-i\frac{N_c^2}{2^4 \pi^2}(\omega_1+\omega_2)-\frac{N_c^2(\ln2 -1)}{2^4\pi^2}(\omega_1^2+\omega_2^2+\omega_1\omega_2)+\cdots,$ 

\noindent and 

$G^{xx-yy|zx|0x}_{\rm AdS}(k_1{=}0,\omega_2{=}0)=\frac{N_c^2}{2^5 \pi^2}\omega_1 k_2+\cdots.$
}.
\section{Acknowledgments}
The authors would like to thank Guy Moore for useful discussions.
This work was supported, in part, by the U.S. Department
of Energy under Grant No.~DE-FG02-97ER41027 and by a
Jeffress research grant, GF12334.

After completing our work we became aware of related work by Omid Saremi and Kiyoumars Sohrabi.

\begin{appendix}

\section{$\eta\tau_\Pi$ contribution to $G^{xy| xz| yz}_{\text{hydro}}$ and $G^{xy| yz| 0x}_{\text{hydro}}$}\label{appendix eta}

In this section we are interested in computing the $\eta\tau_\Pi$ contribution to $T_{xy}$ to second order in gradients, and in the presence of a gravitational perturbation which includes only the shear modes: $h_{x0},h_{xz},h_{y0}$ and $h_{yz}$.
We begin with
\be
(\Pi^{\mu\nu})_{\eta\tau_\Pi}=\eta\tau_\Pi\bigg(
{}^{\langle}U\nabla\sigma^{\mu\nu\rangle}
+\tfrac 13 \nabla\cdot U \,\sigma^{\mu\nu}\bigg)\label{pi eta tau}
\ee
where
\be
{}^{\langle}U\nabla\sigma^{\mu\nu\rangle}=\Delta^{\mu\rho}\Delta^{\nu\sigma}
U^\zeta\nabla_\zeta \sigma_{\rho\sigma}-\tfrac 13 \Delta^{\mu\nu}\Delta^{\rho\sigma} U^\zeta \nabla_\zeta\sigma_{\rho\sigma}\label{u nabla sigma}.
\ee
Given the set of metric perturbations considered, $\sigma^{0\mu}=O(h^2)$. The non-vanishing components of $\sigma^{ij}$, to linear order in the shear metric fluctuations, are $\sigma^{xz}$ and $\sigma^{yz}$ given in (\ref{sigma perp z}) (where we recall that the shear tensor $\sigma$ is symmetric). So for $(\Pi^{xy})_{\eta\tau_\Pi}$ only the first term in (\ref{pi eta tau}) is non-vanishing.

We will also need the projectors $\Delta^{\mu\nu}$ to order $O(h^2)$:
\bea
\Delta^{\mu\nu}=\begin{pmatrix}
0&U_x& U_y&0\\
U_x&1&0&-h_{xz}\\
U_y&0&1&-h_{yz}\\
0&-h_{xz}&-h_{yz}&1
\end{pmatrix}+O(h^2)
\eea
where $U_x, U_y$ are given in (\ref{eq:uperp}).
Given that $\Delta^{xy}$ vanishes to $O(h^2)$, only the first term in (\ref{u nabla sigma}) for $\mu=x, \nu=y$  contributes to terms quadratic in the metric fluctuations:
\be
 \partial_0 \sigma_{xy}^{(2)}-(\Gamma^{z}_{0x}\sigma_{yz}+\Gamma^{z}_{0y}\sigma_{xz})+
(\Delta^{xz}\partial_0 \sigma_{yz}+\Delta^{yz}\partial_0 \sigma_{xz}).
\ee
The last piece we must compute is $\sigma_{xy}$ to second order in the fluctuations 
\be
\sigma_{xy}^{(2)}=h_{x\mu}\sigma^{(1)\,\mu y}+h_{\mu y }\sigma^{(1)\,x\mu}+\sigma^{(2)\, xy}
\ee
  where $\sigma^{(2)\, xy}$ is the truncation of (\ref{shear}) to second order in fluctuations. Using that the second order expansion of the Christoffel symbols is
\be
\Gamma^{(2)}{}_{\mu\nu}^{\lambda}=-g^{\lambda\rho}h_{\rho\sigma}\Gamma^{(1)}{}_{\mu\nu}^{\sigma},
\ee
the shear tensor $\sigma_{xy}^{(2)}$ evaluates to
\be
\sigma_{xy}^{(2)}=\partial_0(U_x U_y)+{\text{other h's}}.
\ee
Substituting everything into (\ref{pi eta tau}) we arrive at
\bea
(\Pi^{xy})_{\eta\tau_\Pi}&=&\eta\tau_\Pi\bigg(\partial_0^2( U_x U_y)-\tfrac 12 (\dot h_{zx}-
h_{0x}')(U_y'-h'_{0y}+\dot h_{zy})-\tfrac 12 (\dot h_{zy}-
h_{0y}')(U_x'-h'_{0x}+\dot h_{zx})\nn\\
&-&h_{xz}(\dot U'_y-\dot h'_{0y}+\ddot h_{zy})
-h_{yz}(\dot U'_x-\dot h'_{0x}+\ddot h_{zx})+{\text{other h's}}\bigg).
\eea
Recalling that $U_\perp$ velocities vanish for $z$-independent metric perturbations (that is, $U_\perp$ vanishes at zero spatial momentum), we can now
extract the desired $\eta\tau_\Pi$ contribution to the correlators
$G_{\text{hydro}}^{xy| xz| yz}(k_1{=}0,k_2{=}0)$ and $G^{xy|yz|0x}_{\text{hydro}}(k_1{=}0,\omega_2{=}0)$:
\bea
G_{\text{hydro}}^{xy| xz| yz}(k_1{=}0,k_2{=}0)=\eta\tau_\Pi(\omega_1\omega_2+\omega_1^2+\omega_2^2)+\dots+\mbox{$\eta\tau_\Pi$-independent terms}\\
G^{xy|yz|0x}_{\text{hydro}}(k_1{=}0,\omega_2{=}0)=\tfrac 12 \eta\tau_\Pi k_2\omega_1+\dots+\mbox{$\eta\tau_\Pi$-independent terms}.
\eea
More general explicit computations show that there are no other second-order hydro coefficients contributions to (\ref{lambda1 4}, \ref{lambda2 3}, \ref{lambda3 2}) besides the ones we explicitly went over in Section 2 and in this appendix.
\section{More Kubo formulae}\label{kubo}
Using the method outlined in Section 2 one can derive a variety of Kubo relations. We give below a few more to supplement (\ref{lambda1 4}), (\ref{lambda2 3}) and (\ref{lambda3 2}). To simplify notation we suppress writing the momentum dependence of a retarded ${\rm raa}$ 3-point correlator $G^{..|..|..}(\bm q=\bm q_1+\bm q_2; -\bm q_1, -\bm q_2)$, with the understanding that the momenta $-\bm q_1, -\bm q_2$ are associated with the second pair and third pair of indices of the 3-point correlator respectively. 
\subsection{$\lambda_1$}
\bea 
&&\lim_{\substack{\omega_1\to0 \\ \omega_2\to 0}}
\partial_{\omega_1}\partial_{\omega_2} \lim_{\substack{k_1\to0 \\ k_2\to 0}}
G_{\text{hydro}}^{xy| xy| xx}=-\tfrac 23 \lambda_1-\tfrac 12 \kappa +\tfrac 53 \eta\tau_\Pi\nn\\
&&\lim_{\substack{\omega_1\to0 \\ \omega_2\to 0}}
\partial_{\omega_1}\partial_{\omega_2} \lim_{\substack{k_1\to0 \\ k_2\to 0}}
G_{\text{hydro}}^{xy| xy| zz}=\tfrac 43 \lambda_1-\tfrac 12 \kappa -\tfrac 13 \eta\tau_\Pi\nn\\
&&\lim_{\substack{\omega_1\to0 \\ \omega_2\to 0}}
\partial_{\omega_1}\partial_{\omega_2} \lim_{\substack{k_1\to0 \\ k_2\to 0}}
G_{\text{hydro}}^{xx| xy| xy}=-\tfrac 23 \lambda_1-\tfrac 16 \kappa + \eta\tau_\Pi\nn\\
&&\lim_{\substack{\omega_1\to0 \\ \omega_2\to 0}}
\partial_{\omega_1}\partial_{\omega_2} \lim_{\substack{k_1\to0 \\ k_2\to 0}}
G_{\text{hydro}}^{zz| xy| xy}=\tfrac 43\lambda_1-\tfrac 16 \kappa -\eta\tau_\Pi\nn\\
&&\lim_{\substack{\omega_1\to0 \\ \omega_2\to 0}}
\partial_{\omega_1}\partial_{\omega_2} \lim_{\substack{k_1\to0 \\ k_2\to 0}}
G_{\text{hydro}}^{yz| xy| xz}=-\lambda_1+\eta\tau_\Pi\label{kubo lambda1}.
\eea
Noting that $O(3)$ rotational symmetry is restored in the limit of vanishing spatial momenta, we can use (\ref{kubo lambda1}) together with $O(3)$ covariance to compute other correlators. E.g. $G^{xx|xy|xy}(k_1{=}0,k_2{=}0)=G^{xx|xz|xz}(k_1{=}0,k_2{=}0)=G^{zz|xz|xz}(k_1{=}0,k_2{=}0)$ and $G^{zz|xy|xy}(k_1{=}0,k_2{=}0)=G^{yy|xz|xz}(k_1{=}0,k_2{=}0)$. Rotational symmetry can also be used to check the last Kubo relation  in (\ref{kubo lambda1}): a $\pi/4$ rotation in the $(y,z)$ plane can be used to show that $G^{yz|xy|xz}(k_1{=}0,k_2{=}0)=\tfrac 12 [G^{yy|xz|xz}(k_1{=}0,k_2{=}0)-G^{yy|xy|xy}(k_1{=}0,k_2{=}0)]$.

\subsection{$\lambda_2$}
The order of limits ($\omega_1\to 0$ and $k_2\to 0$) is important here (due to the presence of a sound pole in the correlators). On the one hand,
\bea
&&\lim_{k_2\to0}\partial_{k_2}\lim_{\omega_1\to 0}
\partial_{\omega_1}\lim_{\substack{k_1\to0 \\ \omega_2\to 0}}
G_{\text{hydro}}^{xx| xz| 0x}=-\lambda_2-\kappa+2\eta\tau_\Pi\nn\\
&&\lim_{k_2\to0}\partial_{k_2}\lim_{\omega_1\to 0}
\partial_{\omega_1}\lim_{\substack{k_1\to0 \\ \omega_2\to 0}}
G_{\text{hydro}}^{yy| xz| 0x}=-\tfrac 12 \lambda_2-\kappa+\eta\tau_\Pi
\eea
but a different order of limits yields
\bea
&&\lim_{\omega_1\to 0}
\partial_{\omega_1}\lim_{k_2\to0}\partial_{k_2}\lim_{\substack{k_1\to0 \\ \omega_2\to 0}}
G_{\text{hydro}}^{xx| xz| 0x}=-\tfrac 12\lambda_2-\tfrac 12\kappa+\tfrac 43\eta\tau_\Pi+\frac{\eta^2}{2\bar\epsilon}
\nn\\
&&\lim_{\omega_1\to 0}
\partial_{\omega_1}\lim_{k_2\to0}\partial_{k_2}\lim_{\substack{k_1\to0 \\ \omega_2\to 0}}
G_{\text{hydro}}^{zz| xz| 0x}=\tfrac 12\lambda_2+\tfrac 12\kappa-\tfrac 23\eta\tau_\Pi-\frac{\eta^2}{\bar\epsilon}\nn\\
&&\lim_{\omega_1\to 0}
\partial_{\omega_1}\lim_{k_2\to0}\partial_{k_2}\lim_{\substack{k_1\to0 \\ \omega_2\to 0}}
G_{\text{hydro}}^{yz| xy| 0x}=\tfrac 14 \lambda_2+\tfrac 12\kappa-\tfrac 12\eta\tau_\Pi-\frac{3\eta^2}{4\bar\epsilon}.
\eea
\subsection{$\lambda_3$}
\bea
&&\lim_{\substack{k_1\to0 \\ k_2\to 0}}
\partial_{k_1}\partial_{k_2} \lim_{\substack{\omega_1\to0 \\ \omega_2\to 0}}
G_{\text{hydro}}^{xx| 0y| 0y}=\tfrac 12\lambda_3,\nn\\
&&\lim_{\substack{k_1\to0 \\ k_2\to 0}}
\partial_{k_1}\partial_{k_2} \lim_{\substack{\omega_1\to0 \\ \omega_2\to 0}}
G_{\text{hydro}}^{00| 0x| 0x}=\tfrac 12\lambda_3.
\eea
In the relation (\ref{lambda2 3}) given in the main text, one gets the same
answer independent of the order of limits $\omega_1\to0$ and $k_2\to 0$.
\section{Retarded bulk-to-boundary propagators}\label{Bb prop}
The metric fluctuations, solutions to the linearized Einstein equations, are in general coupled. We work in the gauge 
\be
\delta g_{ M 5}=0, \qquad M=0,1,2,3,5=t,x,y,z,u.
\ee 
We define
\be
\delta g^{M}_N\equiv \bar g^{MP} \delta g_{PN} 
\ee 
The fluctuations considered in this paper are independent of $x,y$ coordinates.
Relative to the rotation group $SO(2)$ about the $z$ axis, one distinguishes tensor fluctuations ($\delta g_{xy}; \delta g_{xx}-\delta g_{yy}$), vector 
fluctuations ($\delta g_{0t}, \delta g_{xz}$ ; and $\delta g_{y0}, \delta g_{yz}$, and scalar fluctuations ($\delta g_{00},\delta g_{zz},\delta g_{0z},$ and $\delta g_{xx}+\delta g_{yy}$). The tensor fluctuations are completely decoupled, while the vector and scalar fluctuations all mix within their respective sectors for generic $\omega, k$. For the vector, $\delta g_{x0}$ mixes with $\delta g_{xz}$ and $\delta g_{y0}$ mixes with $\delta g_{yz}$. 

Since we are interested in the hydrodynamic regime, where the fields vary slowly with $t, z$, the equations of motion are solved in the bulk perturbatively in $\omega, k$. The bulk-to-boundary causal propagator of coupled functuations will contain one term which behaves like an incoming/outgoing wave at the horizon, in addition to terms which are diffeomorphism terms. The existence of these diffeomorphism terms is inferred by solving the equations of motion near the horizon to leading order, substituting an ansatz of the type $F(u) (1-u)^r$ as in \cite{Policastro:2002tn}. The values of $r=\pm i\omega/2$ correspond to the incoming/outgoing waves; the other possible values of the exponent (e.g. $r =0,-1/2$) correspond to the diffeomorphism terms.  Lastly, we require that the bulk fields approach prescribed values at the $u=0$ boundary: 
$\delta g^N_M\,\stackrel{u\to0}{\longrightarrow}\,h_\mu^\nu$.

We find\footnote{We recall that, for simplicity of notation, we work in units where $2\pi T=1$. Alternatively, one should think of $\omega$ and $k$ as energy and momentum made dimensionless by division with $2\pi T$.}
\bea
\delta g^0_0&=&C_1 (1-u)^{-i\omega/2} \bigg(-\frac{2k^2 (1-u)}3+\dots\bigg)
+D_0 - D_5 \frac{f-uf'+2\omega^2 u}{\sqrt f}\nn\\
\delta g^x_x&=&C_1 (1-u)^{-i\omega/2}\bigg(1-i\frac \omega2 \ln(1+u)+\omega^2(-\frac 12 {\text{Li}}_2(\frac{1-u}{2})+\frac{1}{8}\ln^2(1+u)
\nn\\
&+&\ln(1+u)(1-\frac 12 \ln 2))
+k^2(\frac 13 \ln(1+u)+\frac 23(1-u))+\dots\bigg)\nn\\
&+& C_2 (1-u)^{-i\omega/2}\bigg(1-i\frac \omega2 \ln(1+u) 
+\omega^2(-\frac 12 \text{Li}_2(\frac{1-u}{2})+\frac{1}{8}\ln^2(1+u)
\nn\\
&+&\ln(1+u)(1-\frac 12 \ln 2))
-k^2\ln(1+u)+\dots\bigg)
- D_5 \sqrt{f}\nn\\ 
\delta g^y_y&=&C_1 (1-u)^{-i\omega/2}\bigg(1-i\frac \omega2 \ln(1+u)+\omega^2(-\frac 12 \text{Li}_2(\frac{1-u}{2})+\frac{1}{8}\ln^2(1+u)
\nn\\
&+&\ln(1+u)(1-\frac 12 \ln 2))
+k^2(\frac 13 \ln(1+u)+\frac 23(1-u))+\dots\bigg)\nn\\
&-& C_2 (1-u)^{-i\omega/2}\bigg(1-i\frac \omega2 \ln(1+u) 
+\omega^2(-\frac 12 \text{Li}_2(\frac{1-u}{2})+\frac{1}{8}\ln^2(1+u)
\nn\\
&+&\ln(1+u)(1-\frac 12 \ln 2))
-k^2\ln(1+u)+\dots\bigg)
- D_5 \sqrt{f}\nn\\ 
\delta g^z_z&=&C_1 (1-u)^{-i\omega/2}\bigg(-2+i \omega \ln(1+u)+\omega^2 (\text{Li}_2(\frac{1-u}{2})-\frac 14\ln^2(1+u)\nn\\
&-&(2-\ln 2)\ln(1+u)) 
+k^2(-\frac 23\ln(1+u)+\frac23 (1-u))+\dots\bigg)\nn\\
&-&D_3\frac{2k}{\omega}+D_5(-\sqrt f + 2 k^2 \arcsin(u)) \nn\\ 
\delta g^z_{0}&=&C_1(1-u)^{-i\omega/2}(-i k f +\omega k (1-u)(u+\frac 12(1+u)(\ln(1+u)-2\ln 2)+\dots\bigg)\nn\\
&+&D_3+D_0 \frac{k f}{2\omega} 
-D_5 \omega k(u\sqrt f+\arcsin(u))\nn\\
\delta g^x_0&=&C_3 (1-u)^{-i\omega/2}k\bigg(-i{f}+\frac{\omega}2 f\ln\frac{1+u}2+(\omega+ik^2)u(1-u)+\dots\bigg)
+D_1\frac{\omega}{k}\nn\\
\delta g^y_0&=&C_4 (1-u)^{-i\omega/2}k\bigg(-i{f}+\frac{\omega}2 f\ln\frac{1+u}2+(\omega+ik^2)u(1-u)+\dots\bigg)
+D_2\frac{\omega}{k}\nn\\
\delta g^x_z&=&C_3 (1-u)^{-i\omega/2}\bigg(-2+i\omega\ln\frac{1+u}{2}+k^2+\dots\bigg)
-D_1\nn\\
\delta g^y_z&=&C_4 (1-u)^{-i\omega/2}\bigg(-2+i\omega\ln\frac{1+u}{2}+k^2+\dots\bigg)
-D_2\nn
\eea
\bea
\delta g^x_y&=&C_5 (1-u)^{-i\omega/2}\bigg(1-i\frac\omega 2 \ln({1+u})+\omega^2(-\frac 12
\text{Li}_2(\frac{1-u}2) \nn\\
&+& \frac 18\ln^2(1+u)+(1-\frac {\ln 2}2)\ln(1+u))-k^2\ln(1+u)+\dots\bigg)
\label{bulk metrics}
\eea
where the coefficients $C_1,\dots C_5, D_0, \dots,D_3, D_5$ are given in terms of the boundary fields as follows\footnote{The appearance of higher order sound and diffusion poles is only an artefact of expanding in small $\omega, k$ of simple poles which have additional attenuation pieces. For example, $C_3$ can be repackaged as $k^2(1-(i/2)\omega \ln2+\dots)/(k^2-i\omega(2-k^2)+\dots)$.} :
\bea
C_1&=&\bigg(\frac{k^2}{3\omega^2-k^2}-i\frac{2\omega k^4}{(3\omega^2-k^2)^2}+\frac{k^2A}{24(3\omega^2-k^2)^3}+\dots\bigg)
h^0_{0}\nn\\
&+&\bigg(\frac{\omega^2-k^2}{2(3\omega^2-k^2)}-i\frac{\omega k^2(\omega^2-k^2)}{(3\omega^2-k^2)^2}
+\frac{(\omega^2-k^2)A}{48(3\omega^2-k^2)^3}+\dots\bigg)(h^x_{x}+h^y_{y})\nn\\
&+&\bigg(-\frac{\omega^2}{3\omega^2-k^2}+i\frac{2\omega^3 k^2}{(3\omega^2-k^2)^2}
-\frac{\omega^2 A}{24(3\omega^2-k^2)^3}+\dots\bigg)
h^z_{z}\nn\\
&+&\bigg(-\frac{2\omega k}{3\omega^2-k^2}+i\frac{4\omega^2 k^3}{(3\omega^2-k^2)^2}
-\frac{\omega k A}{12(3\omega^2-k^2)^3}+\dots\bigg)
h^z_{0}
\nn\\
A&=&9\omega^6(\pi^2-6\ln^2{2})-6\omega^4 k^2 (\pi^2+24\ln 2-6\ln^2 2)+
\omega^2 k^4(\pi^2+48\ln 2-6\ln^2 2)-32 k^6\nn\\ 
C_2&=&\bigg(\frac 12 +\frac{\omega^2(\pi^2-6\ln^2{2})}{48}+\dots\bigg)(h^x_{x}-h^y_{y})\nn\\
C_3&=&\bigg(\frac{ik}{2(-i\omega+\tfrac 12 k^2)}-
\frac{\omega k((i\omega+\tfrac 12 k^2)\ln 2-k^2)}{4(-i\omega+\tfrac 12 k^2)^2}+\dots\bigg)(h_0^x+\frac{\omega}{k} h_z^x)\nn\\
C_4&=&\bigg(\frac{ik}{2(-i\omega+\tfrac 12 k^2)}
-\frac{\omega k((i\omega+\tfrac 12 k^2)\ln 2-k^2)}{4(-i\omega+ \tfrac 12k^2)^2}+\dots\bigg)(h_0^y+\frac{\omega}{k} h_z^y)\nn\\
C_5&=&\bigg(1+\frac{\omega^2(\pi^2-6\ln^2{2})}{24}+\dots\bigg)h^x_{y}
\eea
\bea
D_0&=&\bigg(\frac{3\omega^2}{3\omega^2-k^2}-i\frac{2\omega k^4}{(3\omega^2-k^2)^2}+\frac{2\omega^2 k^4(3\omega^2(2-\ln 2)+k^2(\ln 2-4))}{(3\omega^2-k^2)^3}+\dots\bigg)
h^0_{0}\nn\\
&-&\bigg(\frac{\omega^2}{3\omega^2-k^2}+i\frac{\omega k^2(\omega^2-k^2)}{(3\omega^2-k^2)^2}-\frac{\omega^2 k^2(\omega^2-k^2)(3\omega^2(2-\ln 2)+k^2(\ln 2-4))}{(3\omega^2-k^2)^3}+\dots\bigg)\nn\\&\times&(h^x_{x}+h^y_{y})\nn\\
&+&\bigg(-\frac{\omega^2}{3\omega^2-k^2}+i\frac{2\omega^3 k^2}{(3\omega^2-k^2)^2}-\frac{2\omega^4 k^2 (3\omega^2(2-\ln2)+k^2(\ln 2-4))}{(3\omega^2-k^2)^3}+\dots\bigg)h^z_{z}\nn\\
&+&\bigg(-\frac{2\omega k}{3\omega^2-k^2}+i\frac{4\omega^2 k^3}{(3\omega^2-k^2)^2}-\frac{4\omega^3 k^3(3\omega^2(2-\ln2)+k^2(\ln 2-4))}{(3\omega^2-k^2)^3}+\dots\bigg)
h^z_{0}\nn\\
D_1&=&\bigg(-i\frac{k}{-i\omega+\tfrac 12 k^2}+\frac{k^3(\omega\ln 2+(i/2)k^2)}{2(
-i\omega+\tfrac 12 k^2)^2}+\dots\bigg)h^x_0\nn\\
&+&\bigg(-\frac{k^2}{2(-i\omega+\tfrac 12 k^2)}+\frac{k^2\omega(\omega\ln 2+(i/2)k^2)}{2(-i\omega+\tfrac 12 k^2)^2}+\dots\bigg)h^x_z\nn
\eea
\bea
D_2&=&\bigg(-i\frac{k}{-i\omega+\tfrac 12 k^2}+\frac{k^3(\omega\ln 2+(i/2)k^2)}{2(
-i\omega+\tfrac 12 k^2)^2}
+\dots\bigg)h^y_0\nn\\
&+&\bigg(-\frac{k^2}{2(-i\omega+\tfrac 12 k^2)}+\frac{k^2\omega(\omega\ln 2+(i/2)k^2)}{2(-i\omega+\tfrac 12 k^2)^2}+\dots\bigg)h^y_z\nn\\
D_3&=&\bigg(-\frac{3\omega k}{2(3\omega^2-k^2)}+
i\frac{3\omega^2 k^3}{(3\omega^2-k^2)^2}+\frac{\omega k^3 (9\omega^4\ln 2-3 \omega^2 k^2\ln2 +2 k^4)}{(3\omega^2-k^2)^3}+\dots\bigg)h^0_{0}\nn\\
&+&\bigg(\frac{\omega k}{2(3\omega^2-k^2)}+i
\frac{3\omega^2 k (\omega^2-k^2)}{2(3\omega^2-k^2)^2}
+\frac{\omega k(\omega^2-k^2)(9\omega^4\ln 2-3 \omega^2 k^2\ln2 +2 k^4)}{2(3\omega^2-k^2)^3}+\dots\bigg)\nn\\&\times&(h^x_{x}+h^y_{y})\nn\\
&+&\bigg(\frac{\omega k}{2(3\omega^2-k^2)}-i
\frac{3\omega^4 k}{(3\omega^2-k^2)^2}
-\frac{\omega^3 k(9\omega^4\ln 2-3 \omega^2 k^2\ln2 +2 k^4)}{(3\omega^2-k^2)^3}+\dots\bigg)h^z_{z}\nn\\
&+&\bigg(\frac{3\omega^2}{3\omega^2-k^2}-i
\frac{6\omega^3 k^2}{(3\omega^2-k^2)^2}
-\frac{2\omega^2 k^2(9\omega^4\ln 2-3 \omega^2 k^2\ln2 +2 k^4)}{(3\omega^2-k^2)^3}+\dots\bigg)h^3_{0}\nn
\eea
\bea
D_5&=&\bigg(\frac{k^2}{3\omega^2-k^2}-i\frac{2\omega k^4}{(3\omega^2-k^2)^2}+
\frac{2k^4(9\omega^4(1-\ln2)+3\omega^2 k^2(\ln 2-2)-k^4)}{3(3\omega^2-k^2)^3}+\dots\bigg)h^0_{0}\nn\\
&-&\bigg(\frac{\omega^2}{3\omega^2-k^2}+i\frac{\omega k^2(\omega^2-k^2)}{(3\omega^2-k^2)^2}-
\frac{k^2(\omega^2-k^2)(9\omega^4(1-\ln2)+3\omega^2 k^2(\ln 2-2)-k^4)}{3(3\omega^2-k^2)^3}+\dots\bigg)\nn\\
&\times&(h^x_{x}+h^y_{y})\nn\\
&+&\bigg(-\frac{\omega^2}{3\omega^2-k^2}+i\frac{2\omega^3 k^2}{(3\omega^2-k^2)^2}-
\frac{2\omega^2 k^2(9\omega^4(1-\ln2)+3\omega^2 k^2(\ln 2-2)-k^4)}{3(3\omega^2-k^2)^3}+\dots\bigg)h^z_{z}\nn\\
&+&\bigg(-\frac{2\omega k}{3\omega^2-k^2}+i\frac{4\omega^2 k^3}{(3\omega^2-k^2)^2}-
\frac{4\omega k^3(9\omega^4(1-\ln2)+3\omega^2 k^2(\ln 2-2)-k^4)}{3(3\omega^2-k^2)^3}+\dots\bigg)h^z_{0}\nn\\
\eea
The diffeomorphism parameters which give rise to the diffeomorphism terms in the bulk-to-boundary propagators are:
\bea
\xi_M(\omega, k,u)&=&\bigg(-i \frac{D_0f}{2\omega u}+iD_5 \omega \sqrt{f},
i\frac{D_1}{k u},
i\frac{D_2}{k u},
i \frac{D_3}{\omega u}-i\frac{D_5k\text{arcsin}(u)}{u},
\frac{D_5}{u\sqrt{f}}\bigg)
\nn\\
\xi^M(\omega, k,u)&=&\bigg(i\frac{D_0}{2\omega}
-i D_5\frac{\omega u}{ \sqrt{f}},
i\frac{D_1}{k},
i\frac{D_2}{k},
i\frac{D_3}{\omega}-i D_5 k\text{arcsin}(u),D_5\sqrt{f}u\bigg)\label{xi}
\eea
That is, the ``$D$'' terms in (\ref{bulk metrics}) are given by 
$(\delta g^M_N)_{\rm {diff}}=\nabla_N\xi^M+\nabla^M\xi_N$. 
The $D_0, D_1, D_2, D_3$ terms in (\ref{xi}) generate boundary diffeomorphisms. The $D_5$ term generates an infinitesimal scale transformation in the vicinity of the $u=0$ boundary, plus additional boundary diffeomorphisms to preserve the gauge condition $\delta g_{M 5}=0$.
Note that these diffeomorphism parameters induce singular gauge transformations at the horizon. This is a consequence of working in the gauge $\delta g_{M 5}=0$.

\section{Sound mode 2-point correlators}\label{sound}
From the coupled fluctuations $\delta g^x_{x},\delta g^y_{y}, \delta g^x_{z},\delta g^z_{z},\delta g^0_{0}$ (where for simplicity of notation we lumped together the sound modes with one more fluctuation, namely $\delta g^x_{x}-\delta g^y_{y}$ which can actually be decoupled) and from the form of the on-shell
quadratic gravitational action (see Appendix E) we get
\bea
G^{xx| xx}_{\rm AdS}&=&-4\frac{\delta^2{\cal S}}{\delta h_{xx}\delta h_{xx}}\nn\\
&=&\frac{N_c^2}{2^4\pi^2}\bigg[\frac{(7\omega^2-k^2) }{2^3(3\omega^2-k^2)}
-i\frac{(3\omega^4-3\omega^2 k^2+k^4)\omega }{(3\omega^2-k^2)^2}
\nn\\
&+&\frac{1}{2(3\omega^2-k^2)^3}
\bigg(\omega^2(18\omega^6(1-\ln2)
+\omega^4 k^2 (24\ln 2-39)+\omega^2 k^4(28-12\ln 2)\nn\\
&+&k^6(2\ln 2-7))\bigg)
+\dots\bigg]
\nn\\ 
G^{xx| yy}_{\rm AdS}&=&-4\frac{\delta^2{\cal S}}{\delta h_{xx}\delta h_{yy}}\nn\\
&=&\frac{N_c^2}{2^4\pi^2}\bigg[\frac{(\omega^2+k^2)}{2^3(3\omega^2-k^2)}
+i\frac{(3\omega^4-k^4)\omega }{2(3\omega^2-k^2)^2}\nn\\ 
&+&\frac{1}{2(3\omega^2-k^2)^3}\bigg(9\omega^8(\ln 2-1)+
\omega^6 k^2 (15-3\ln 2)-\omega^4 k^4 (8+3\ln 2)\nn\\
&+&(3+\ln 2) \omega^2 k^6 -k^8)\bigg)+\dots\bigg]\nn \\ 
G^{00,00}_{\rm AdS}&=& -4\frac{\delta^2{\cal S}}{\delta^2 h_{00}}\nn\\
&=&\frac{N_c^2}{2^4\pi^2}\bigg[
\frac{3(5 k^2-3\omega^2) }{2^3(3\omega^2-k^2)}
-i\frac{3 \omega k^4 }{(3\omega^2-k^2)^2}
\nn\\&+&\frac{k^4(9\omega^4(1-\ln 2) + 3 \omega^2 k^2 (\ln 2-2) -k^4)}{(3\omega^2-k^2)^3}+\dots\bigg]\nn
\eea
\bea
G^{zz| zz}_{\rm AdS}&=&-4\frac{\delta^2{\cal S}}{\delta  h_{zz}\delta  h_{zz}}\nn\\
&=&\frac{N_c^2}{2^4\pi^2}\bigg[
\frac{(7\omega^2-k^2)}{2^3(3\omega^2-k^2)}
-i\frac{3\omega^5}{(3\omega^2-k^2)^2}\nn\\
&+&\frac{\omega^4(9\omega^4(1-\ln 2)+3\omega^2 k^2 (\ln 2-2)-k^4)\pi^2 }{(3\omega^2-k^2)^3}+\dots\bigg]\nn\\ 
G^{00| 0z}_{\rm AdS}&=&-2\frac{\delta^2{\cal S}}{\delta h_{00} \delta h_{0z}}\nn\\
&=&\frac{N_c^2}{2^4\pi^2}\bigg[\frac{3\omega k}{2(3\omega^2-k^2)}\nn\\
&-&i\frac{3 \omega^2 k^3}{(3\omega^2-k^2)^2}
+\frac{\omega k^3(9\omega^4(1-\ln 2)+3\omega^2 k^2 (\ln 2-2)-k^4) }{(3\omega^2-k^2)^3}+\dots\bigg]\nn
\eea
\bea
G^{00| zz}_{\rm AdS}&=&-4\frac{\delta^2{\cal S}}{\delta h_{00} \delta h_{zz}}\nn\\
&=&\frac{N_c^2}{2^4\pi^2}\bigg[\frac{3(\omega^2+k^2)}{2^3(3\omega^2-k^2)}
-i\frac{3\omega^3 k^2 }{(3\omega^2-k^2)^2}
\nn\\&+&\frac{\omega^2 k^2 (9\omega^4(1-\ln2) + 3\omega^2 k^2(\ln 2-2)-k^4)}{(3\omega^2-k^2)^3}+\dots\bigg]\nn\\ 
G^{0z| 0z}_{\rm AdS}&=&-\frac{\delta^2{\cal S}}{\delta h_{0z}\delta h_{0z}}\nn\\
&=&\frac{N_c^2}{2^4\pi^2}\bigg[
\frac{(9\omega^2+k^2)}{2^3(3\omega^2-k^2)}
-i\frac{3\omega^3 k^2}{(3\omega^2-k^2)^2}
\nn\\&+&\frac{\omega^2 k^2 (9\omega^4(1-\ln2) + 3\omega^2 k^2(\ln 2-2)-k^4)}{(3\omega^2-k^2)^3}+\dots\bigg]\nn\\ 
G^{0z| zz}_{\rm AdS}&=&-2\frac{\delta^2{\cal S}}{\delta h_{0z}\delta h_{zz}}\nn\\
&=&\frac{N_c^2}{2^4\pi^2}\bigg[\frac{\omega k}{2(3\omega^2-k^2)}
-i\frac{3 \omega^4k }{(3\omega^2-k^2)^2}\nn\\
&+&\frac{k\omega^3(9\omega^4(1-\ln 2)+3(\ln 2-2)\omega^2 k^2-k^4)}{(3\omega^2-k^2)^3}+\dots\bigg]\nn\\ 
G^{00| xx}_{\rm AdS}&=&-4\frac{\delta^2{\cal S}}{\delta h_{00} \delta h_{xx}}\nn\\
&=&\frac{N_c^2}{2^4\pi^2}\bigg[\frac{3(\omega^2+k^2)}{2^3(3\omega^2-k^2)}
+i\frac{3\omega   k^2(\omega^2-k^2) }{2(3\omega^2-k^2)^2}\nn\\
&-&\frac{k^2(\omega^2-k^2)(9\omega^4(1-\ln 2)+3\omega^2 k^2 (\ln2-2)-k^4)}{2(3\omega^2-k^2)^3}+\dots\bigg]\nn
\eea
\bea
G^{0z| xx}_{\rm AdS}&=&-2\frac{\delta^2{\cal S}}{\delta h_{0z}\delta h_{xx}}\nn\\
&=&\frac{N_c^2}{2^4\pi^2}\bigg[\frac{\omega k }{2(3\omega^2-k^2)}
+i\frac{3(\omega^2-k^2) k \omega^2 }{2(3\omega^2-k^2)^2}
\nn\\
&-&\frac{\omega k (\omega^2-k^2)(9\omega^4(1-\ln 2)+3(\ln 2-2)\omega^2 k^2-k^4)}{2(3\omega^2-k^2)^3}+\dots\bigg]\nn\\ 
G^{zz| xx}_{\rm AdS}&=&-4\frac{\delta^2{\cal S}}{\delta h_{zz}\delta h_{xx}}\nn\\
&=&\frac{N_c^2}{2^4\pi^2}\bigg[\frac{(\omega^2+k^2) }{2^3(3\omega^2-k^2)}
+i\frac{3(\omega^2-k^2)\omega^3  }{2(3\omega^2-k^2)^2}\nn\\
&+&\frac{\omega^2  (k^2-\omega^2)(9\omega^4(1-\ln 2)+3(\ln 2-2) \omega^2 k^2 -k^4)}{2(3\omega^2-k^2)^3}+\dots\bigg]\nn\\ 
\eea
These 2-point functions exhibit, as expected, the sound pole\footnote{The sound pole complete expression is \cite{Policastro:2002se}
$v_s k -\frac{i}{2(\epsilon+P)}(\zeta+\frac 43 \eta)k^2$ where $\zeta$ is the bulk viscosity, $\eta$ is the shear viscosity, and the speed of sound is $v_s=\sqrt{\partial P/\partial\epsilon}.$
} expanded in small $\omega, k$:
$\omega\simeq \pm k/\sqrt 3-i\frac {\eta k^2}{6P}$.

\section{On-shell gravitational action}\label{1pt}
For completeness, in this appendix we give the linear, quadratic and cubic gravitational vertex.
This corresponds to expanding  the gravitational action (\ref{act2})
in terms linear, quadratic and cubic in the linearized on-shell fluctuations:
\be
{\cal S}=\bar {\cal S} + \delta^{(1)}{\cal S}+
\delta^{(2)}{\cal S}+\delta^{(3)}{\cal S}+\cdots
\ee

\subsection{First order action}

To linear order in the fluctuations (\ref{act2}) is a boundary term\footnote{The second counterterm, proportional to the boundary Ricci scalar, does not contribute to this order.}:
\bea
&&\delta^{(1)}\bigg(\int_{\cal M} \sqrt {-g}(R-2\Lambda)+\int_{\partial {\cal M}}
\sqrt{-g_{\text{bdy}}}(a+2 K)\bigg)=\frac 12 \int_{\partial {\cal M}}
\sqrt{-g_{\text{bdy}}} \bigg[n^5\bigg(-(\partial_5 g_{\mu\nu})\delta g^{\mu\nu}\nn\\&&\qquad\qquad \qquad\qquad\qquad \qquad\qquad\qquad \qquad\qquad \qquad
 + (\partial_5 g_{\mu\nu}) g^{\mu\nu} \delta g_\rho^\rho\bigg) 
+ a \delta g_\rho^\rho
\bigg]\label{delta1s}
\eea
where $\mu,\nu, \rho=0,1,2,3$, and all indices are raised and lowered with the backgroud metric, e.g. $\delta g^{\mu\nu}=\delta g_{\rho\sigma }\bar g^{\mu\rho} \bar g^{\nu\sigma}$. As expected, the Gibbons-Hawking term  has contributed to the cancellation of the terms linear in derivatives of the metric fluctuations at the boundary. Also, the leading order divergence (proportional to $1/u_B^2$, where the boundary value is of the radial coordinate $u_B \to 0$) from the Einstein-Hilbert plus cosmological constant action and from the Gibbons-Hawking term is canceled by the boundary volume counterterm.

In the black hole background, the on-shell action linear in the boundary fields is finite \cite{Policastro:2002tn}\footnote{In AdS background, as opposed to the AdS-Schwarzschild case we consider, the on-shell action linear in fluctuations is zero.}:
\be
\delta^{(1)}{\cal S}=\frac{N_c^2}{2^6 \pi^2} \int_{u=0} (-\frac 34 \delta g_0^0 + \frac 14\sum_{i=1,2,3}\delta g_i^i)
\ee
where we have introduced the shorthand notation
\be
\int_{u=0}\cdots =\int d^4 {\bm x} \cdots\bigg|_{u=0}.
\ee

For example, the one-point function of the stress tensor $T^{00}$ is obtained by differentiating with respect to $h_0^0\equiv \delta g_0^0|_{u=0}$:
\be
\langle T_0^0\rangle_0=2\frac{\delta{\cal S}}{\delta h_0^0}\bigg|_{h_\mu^\nu=0}=2\frac{\delta (\delta^{(1)}{\cal S})}{\delta h_0^0 }=-\frac{3N_c^2}{2^7 \pi^2}.
\ee
To get the correct dimension for the stress tensor, we can restore the temperature dependence 
\be
\langle T_0^0\rangle_0=(2\pi T)^4 \times \bigg(-\frac{3N_c^2}{2^7\pi^2}\bigg)=-\frac 38 \pi^2 N_c^2 T^4.
\ee

\subsection{Second order action}

Now expand (\ref{act2}) to second order in the linearized on-shell fluctuations. To highlight the role of the second counterterm in (\ref{act2}), and explain why one needs to go beyond the quadratic action given by \cite{Policastro:2002tn}, we first expand
\be
\int_{\cal M} \sqrt {-g}(R-2\Lambda)+\int_{\partial {\cal M}}
\sqrt{-g_{\text{bdy}}}(a+2 K) \label{act1}
\ee 
and collect the $(\delta g)^2$ terms. Since the fluctuations obey the linearized equations of motion, we are left again with a total derivative term (basically the first-order expansion of (\ref{delta1s})). (The bulk term cancelled in the first-order expansion of the action (\ref{act1}) because the background solved the Einstein equations. The bulk term cancels now because the fluctuations are on-shell.) After these considerations, this is what is left:
\bea
&&\frac 14 \int_{u=0} \sqrt{-g_{\text{bdy}}} n^5 \bigg(-(\partial_5 \delta g^\mu_\nu)\delta g^\nu_\mu +( \partial_5 \delta g^\mu_\mu)\delta g^\nu_\nu\bigg)\nn\\
&&+
\frac 14 \int_{u=0}\sqrt{-g_{\text{bdy}}} \bigg[ n^5\bigg(
(\partial_5 g_{\mu\nu})\delta g^\nu_\rho \delta g^{\mu\rho}-(\partial_5 g_{\mu\nu}) g^{\mu\nu} \delta g^\rho_\sigma \delta g^\sigma_\rho-\frac 12 (\partial_5 g_{\mu\nu})\delta g^{\mu\nu}\delta g^\rho_\rho \nn\\
&&\qquad\qquad+
\frac 12 (\partial_5 g_{\mu\nu}) g^{\mu\nu} \delta g_\rho^\rho \delta g_\sigma^\sigma \bigg)
+\frac  12 a \delta g^\mu_\mu \delta g^\nu_\nu - a\delta g^\mu_\nu \delta g^\nu_\mu\bigg]\label{delta2s}.
\eea
Again, the boundary volume counterterm, with its $a$ coefficient, removes a $1/u^2$ divergence from the contact terms in (\ref{delta2s}), besides contributing to the finite terms. Upon substituting the black hole background metric
(\ref{bh metric}),  and 
\be
n^5=-u \sqrt{1-u^2}, \qquad \sqrt{-g_{\text{bdy}}}=\frac{\sqrt{1-u^2}}{u^2}
\label{unit n}
\ee 
in the second order expansion  (\ref{delta2s}) we arrive at
\bea
&&\;\;\frac 18 \int_{u=0}\frac{1}{u}
\partial_5\bigg(-\delta g^\mu_\mu \delta g^\nu_\nu+\delta g^\mu_\nu \delta g^\nu_\mu \bigg)
\nn\\
&&+\frac 14\int_{u=0}\bigg( \frac 34 (h^0_0)^2
-\frac 12 h^0_0 h^i_i + h^0_i h^i_0 + \frac 14 h^i_i h^j_j - \frac 12 h^i_j h^j_i\bigg)\label{delta2s1}
\eea
where $\mu,\nu,\rho=0,1,2,3$; $i,j=1,2,3$ and Einstein summation convention was used\footnote{This agrees with (3.15) in \cite{Policastro:2002tn}. An apparent discrepancy is resolved by noting that the derivative terms in \cite{Policastro:2002tn}  include $\partial_5( (h^3_0)^2 )$, whereas we have $\partial_5 (h_3^0 h_0^3)$. This difference is reflected in a different coefficient of $h^0_3 h^3_0$.}.

However, while the action (\ref{act1}) is properly regularized and gives correct answers for the field theory stress tensor two-point functions up to terms linear in $\omega, k$ (and therefore suffices as long as one is interested in linear hydrodynamic coefficients), it diverges as $1/u$ to quadratic order in $\omega, k$. To remove this divergence a second counterterm is needed, proportional to the boundary Einstein-Hilbert action \cite{Balasubramanian:1999re, Kraus:1999di, Emparan:1999pm, de Haro:2000xn}:
\be
-\frac{R_{\rm AdS}}{d-2}\int_{\partial {\cal M}}\sqrt{-g_{\text{bdy}}} R_{\text{bdy}}
\ee 
where we recall that for us $d=4$ and we set $R_{\rm AdS}=2$. It is easy to see that this second counterterm contributes to leading order only to terms quadratic in $\omega, k$, 
\bea
\frac 12\int_{u=0}\frac{\sqrt{1-u^2}}{u^2}\bigg(\partial^2 \delta g^\mu_\nu+
\partial^\mu \partial_\nu \delta g^\rho_\rho -\partial_\rho\partial^\mu \delta g^\rho_\nu
-\partial^\rho\partial_\nu \delta g_\rho^\mu
-\delta^\mu_\nu(\partial^2 \delta g^\rho_\rho-\partial^\rho \partial_\sigma \delta g^{\sigma}_\rho)\bigg)\delta g_\mu^{\nu}
\label{intermediate1}
\eea
since the curvature tensor is quadratic in derivatives.
In (\ref{intermediate1}) indices are raised and lowered on the partial derivatives with the boundary metric $\bar g_{\text {bdy}\,\mu\nu}=\text{diag}(-f/u,1/u,1/u,1/u)$. 
Since there is one inverse metric per term in (\ref{intermediate1}) we see that the contribution coming from this counterterm will be divergent as $1/u$. The conclusion is that this counterterm's job, to second order in fluctuations and second order in $\omega, k$, is only to remove divergences from (\ref{delta2s1}), without any finite term subtraction.  At fourth order in $\omega, k$ we get a finite contribution, coming from terms linear in $u$ in $\delta g^i_j$. For completeness, we give below the leading order in $\omega, k$ of (\ref{intermediate1}):
\bea
&&\;\;\int_{u=0}\frac{1}{u}\bigg( (\omega^2-k^2)
[(h^1_2)^2-h^1_1 h^2_2]
+(k^2 h^0_0 -h^3_3 \omega^2) (h^1_1 + h^2_2)\nn\\
&&\qquad\qquad
-\omega k h^3_0(h^1_1+h^2_2)+\omega k(h^1_0 h^1_3+h^2_0 h^2_3)
+\omega^2 [(h^1_3)^2+(h^2_3)^2]+k^2[(h^1_0)^2+(h^2_0)^2]\bigg)\nn\\\label{delta2s2}.
\eea
 The second-order on-shell gravitational action $\delta^{(2)}{\cal S}$ is given by the sum of equations (\ref{delta2s1}) and (\ref{delta2s2}) multiplied by the gravitational prefactor $N_c^2/(2^6\pi^2)$.

\subsection{Cubic action}

In writing the cubic action, it is helpful to start with the second-order Einstein-Hilbert plus cosmological constant action
\bea
\delta^{(2)}{\cal S}_{\rm EH}&\equiv&\delta^{(2)}\int_{\cal M}\sqrt{-g}(R-2\Lambda)\nn\\&=&
\frac 12\int_{\cal M}\sqrt{-\bar g}\nabla_M 
\bigg[\nabla^M(\tfrac 34 \delta g^{KL}\delta g_{KL}-\tfrac 14 \delta g^K_K\delta g^L_L)+\tfrac 12 \nabla_N(\delta g^K_K \delta g^{MN})\nn\\
&&\qquad\qquad-
2\nabla_N(\delta g^{MP}\delta g^N_P)
+\delta g^{KN}\nabla_N
\delta g^{M}_K+\delta g^{MN}\nabla_N \delta g^K_K\bigg]\label{delta22}
\eea
where the indices are being raised and lowered with the background metric $\bar g_{MN}$, and the derivatives are background covariant. In general, there is another contribution to the second-order action, which is proportional to the linearized equation of motion. Since $\delta g_{MN}$ solves the linearized equation of motion, the only non-vanishing contribution is the total derivative given in (\ref{delta22}).
In expanding to terms cubic in the fluctuations, the Einstein-Hilbert plus cosmological constant action receives contributions from $\delta^{(1)}(\delta^{(2)}{\cal S}_{\rm EH})$ 
as well as from the expansion of the term proportional to the linearized equations of motion:
\be
\delta^{(3)}{\cal S}_{\rm EH}=\tfrac 13 \delta^{(1)}\bigg(\delta^{(2)}{\cal S}_{\rm EH}\bigg)
+\tfrac 1{3!} \delta^{(1)}\int_{\cal M}\sqrt{-g}\tfrac 12(L_{MN}-\tfrac 12 g_{MN}L)\delta g^{MN}\label{cubic action}.
\ee
A brute force calculation gives
\bea
&&\tfrac 13 \delta^{(1)}\bigg(\delta^{(2)}{\cal S}_{\rm EH}\bigg)
=\int_{\cal M}\sqrt{-\bar g}\nabla_M \bigg[-\tfrac 1{24}(\delta g^K_K)^2
\nabla^M \delta g^L_L+\tfrac 18 \delta g^{MN}\nabla_N (\delta g^K_K)^2
+\tfrac 1{24}(\delta g^K_K)^2 \nabla_N\delta g^{MN}\nn\\
&&\qquad\qquad\qquad\qquad+\tfrac 18 \delta g^K_K\nabla^M(\delta g_{NP}\delta g^{NP})
+\tfrac 1{12}\delta g_{NP}\delta g^{NP} \nabla^M \delta g^K_K
-\tfrac 23 \delta g^{MP}\delta g_{P}^N\nabla_N\delta g^K_K\nn\\
&&\qquad\qquad\qquad\qquad-\tfrac 16 \delta g^K_K\delta g^{NP}\nabla_N\delta g^M_P
-\tfrac 13 \delta g^K_K\delta g^{MP}\nabla_N\delta g^{N}_P
+\tfrac 13 \delta g^{NP}\delta g_{PQ}\nabla_N\delta g^{MQ}\nn\\
&&\qquad\qquad\qquad\qquad+\delta g^{MP}\delta g_{PQ}\nabla_N\delta g^{NQ}
-\tfrac 23 \delta g^{MP}\delta g^{NQ}\nabla_N\delta g_{PQ}
-\tfrac 1{12}\delta g^{PQ}\delta g_{PQ}\nabla_N\delta g^{MN}\nn\\
&&\qquad\qquad\qquad\qquad-\tfrac 13 \delta g^{MN}\nabla_{N}(\delta g^{PQ}\delta g_{PQ})
-\tfrac 23 \delta g^{NP}\delta g_{P}^Q\nabla^M\delta g_{NQ}\bigg]\label{cubic tot deriv}.
\eea
The $L_{MN}$ tensor which appears in (\ref{cubic action}) was introduced in \cite{Arutyunov:1999nw}\footnote{$L_{MN}$ given in equation (2.6) in \cite{Arutyunov:1999nw} has been 
further simplified by commuting two covariant derivatives, and subsequently using that the AdS curvature tensor is $R_{MPN}{}^Q=-(g_{MN} \delta_P^Q - g_{PN}\delta_M^Q)R_{\rm AdS}^{-2}$. Arutyunov and Frolov have further set 
$\rm R_{AdS}=1.$}:
\bea
-\tfrac 12 L_{MN}&\equiv&\delta^{(1)}R_{MN}-\tfrac 12 \delta g_{MN}(\bar R-2\Lambda)
\nn\\
&=&-\tfrac 12(\nabla^K\nabla_K\delta g_{MN}+\nabla_M\nabla_N\delta g^K_K
-\nabla^K\nabla_M\delta g_{NK}-\nabla^K\nabla_N\delta g_{MK})+d\delta g_{MN}
R_{\rm AdS}^{-2}\nn\\
\eea
where we recall that the background Ricci scalar is 
$\bar R=2(d+1)\Lambda/(d-1)=-d(d+1)R_{\rm AdS}^{-2}$ and $d=4$ for us. In terms of $L_{MN}$, the 
linearized equation of motion is $L_{MN}-\tfrac 12 \bar g_{MN} L=0$.
 
In \cite{Arutyunov:1999nw} it was also shown that the second order variation of the equation of motion, 
$\tfrac 12\delta ^{(1)}(L_{MN}-\tfrac 12 \bar g_{MN}L)$  is expressed in terms of another tensor, $V_{MN}$:
\bea
-\tfrac 12\delta^{(1)}L_{MN}&\equiv& V_{MN}=\delta g^{PQ}(\nabla_P\nabla_Q\delta g_{MN}+
\nabla_M\nabla_N\delta g_{PQ}-\nabla_P\nabla_M\delta g_{NQ}
-\nabla_P\nabla_N\delta g_{MQ})\nn\\
&&\qquad\qquad-
\frac 12 (2\nabla_P\delta g^{PQ}-\nabla^Q\delta g^P_P)(\nabla_M\delta g_{NQ}+
\nabla_N\delta g_{MQ}-\nabla_Q\delta g_{MN})\nn\\
&&\qquad\qquad+
\frac 12 [(\nabla_M \delta g^{PQ})\nabla_N\delta g_{PQ}
+2(\nabla^P\delta g^Q_M )\nabla_P\delta g_{QN}
-2(\nabla^P\delta g^Q_M) \nabla_Q\delta g_{PN}].\nn\\
\eea
Lastly, substituting into the second term of the cubic action (\ref{cubic action}), one finds \cite{Arutyunov:1999nw}
\bea
&&\tfrac 1{3!} \delta^{(1)}\int_{\cal M}\sqrt{-g}\tfrac 12(L_{MN}-\tfrac 12 g_{MN}L)\delta g^{MN}
=-\tfrac 16 \int_{\cal M}\sqrt{-\bar g}
(V_{MN}-\tfrac 12 \bar g_{MN}V)
\delta g^{MN}.
\eea
The total derivative terms omitted from the cubic action in \cite{Arutyunov:1999nw} are those we give in (\ref{cubic tot deriv}). We do not throw away these terms, and keep their contribution (from both boundary and horizon) to the retarded momentum space 3-point stress tensor correlators.

For the correlators we evaluate in Section 3.2, we only need cubic vertices of one tensor mode and two shear modes. This cuts down significantly the number of relevant terms from the cubic action (\ref{cubic action}).

The Gibbons-Hawking term contribution to the cubic action is most easily evaluated by making explicit use of the gauge condition $\delta g_{M5}=0$ and of the form of the unit normal vector to the boundary (\ref{unit n}):
\bea
&&\delta^{(3)}{\cal S}_{\rm GH}=\tfrac 13 \delta^{(1)} (\delta^{(2)} {\cal S}_{\rm GH})
\nn\\
&&=\tfrac 13 \delta^{(1)}\bigg[\int_{u=0}
\sqrt{-g}\bar g^{55} \bigg(
(\delta g^{\mu\nu} - \tfrac 12 g^{\mu\nu} \delta g^\rho_\rho)\partial_5 \delta g_{\mu\nu}-(\delta g^\mu_\rho\delta g^{\nu\rho}-\tfrac 12 \delta g^\rho_\rho \delta g^{\mu\nu}) \partial_5 g_{\mu\nu}\nn\\&&
\qquad\qquad
+\tfrac 14(\delta g^{\mu\nu}\delta g_{\mu\nu}-\tfrac 12 (\delta g^\mu_\mu)^2)g^{\rho\sigma}\partial_5 g_{\rho\sigma}\bigg)\bigg]\nn\\
&&=
\int_{u=0}
\sqrt{-\bar g}\bar g^{55}\bigg[-\bigg(
\delta g^\mu_\rho\delta g^{\nu\rho}-
\tfrac 12 \delta g^{\mu\nu}\delta g^\rho_\rho -
\tfrac 14 \bar g^{\mu\nu}(\delta g^{\rho\sigma}\delta g_{\rho\sigma}-
\tfrac 12 (\delta g^\rho_\rho)^2)\bigg)\partial_5\delta g_{\mu\nu}\nn\\
&&\qquad\qquad
+\bigg(\delta g^{\mu\rho}\delta g_{\rho\sigma}\delta g^{\nu\sigma}-\tfrac 12
\delta g^\rho_\rho \delta g^{\mu\sigma}\delta g^{\nu}_{\sigma}
-\tfrac 14\delta g^{\mu\nu}( \delta g^{\rho\sigma}\delta g_{\rho\sigma}-\tfrac 12(\delta g^\rho_\rho)^2)\bigg)\partial_5\bar g_{\mu\nu}\nn\\
&&\qquad\qquad
-\tfrac 16 \bigg(\delta g^{\mu\nu}\delta g_{\mu\rho}\delta g^{\rho}_\nu
-\tfrac 34 \delta g^{\mu\nu}\delta g_{\mu\nu}\delta g^\rho_\rho
+\tfrac 18(\delta g^\rho_\rho)^3\bigg)\bar g^{\zeta\tau}\partial_5\bar g_{\zeta\tau}\bigg].
\eea 
The cubic action $\delta^{(3)}{\cal S}$ is given by the sum of $\delta^{(3)}{\cal S}_{\rm EH}+ \delta^{(3)}{\cal S}_{\rm GH}$ and of the cubic expansion of the boundary counterterms, multiplied by the gravitational prefactor $N_c^2/(2^6\pi^2)$.

\end{appendix}

\end{document}